\documentclass[12pt]{article}
\pdfoutput=1

\usepackage{putex}

\usepackage{geometry}

\usepackage{graphicx} 
\usepackage{amsfonts,amsmath,amssymb,epsf}
\usepackage{graphics} 
\usepackage{physics}
\usepackage{cancel}
\usepackage{tikz}
\usepackage[T1]{fontenc}
\usepackage[utf8]{inputenc}
\usepackage{enumerate}
\usepackage{array}
\usepackage{hyperref}
\usepackage{graphicx}
\usepackage[export]{adjustbox}
\usepackage{subcaption}
\usepackage{cite}
\usepackage{babel}

\usepackage{titlesec}
\titleformat*{\subsection}{\bfseries\boldmath}
\titleformat*{\section}{\bfseries\boldmath}

\newcommand{\wt}{\widetilde}

\usetikzlibrary{arrows.meta,decorations.markings}

\newcommand\be{\begin{equation}}
\newcommand\ee{\end{equation}}

\newcommand{\cC}{{\mathcal C}}

\newcommand{\cka}{\checkmark}
\newcommand{\ckb}{\checkmark\!\!\!\checkmark}
\newcommand{\lphi}{\lambda\psi}
\newcommand{\ot}{\otimes}

\newcommand{\ov}{\, \overline }
\newcommand{\F}[3]{{\rm F}^{(#1)#2}_{#3} }
\newcommand{\G}[3]{{\rm G}^{(#1)#2}_{#3} }
\newcommand{\R}[2]{{\rm R}^{(#1)#2}}

\newcommand{\myarrow}[1][.6]{postaction={decorate,decoration={
                markings,
                mark=at position .6 with {\arrow{Stealth}}}}}

\def\comma{\,,}
\def\period{\,.}

\newcommand{\cD}{\mathcal{D}}

\newcommand*\circled[1]{\tikz[baseline=(char.base)]{
        \node[shape=circle,draw,scale=0.95,inner sep=2pt] (char) {#1};}}

\usepackage{orcidlink}

\newcommand\doi[2]        {\href{https://dx.doi.org/#1}{#2}}

\numberwithin{equation}{section}

\date{}

\begin{document}
\preprint{DESY-25-053, HBzM 989, ZMP-HH/25-6}

	\institution{DESY}{$^{(1,2)}$ Deutsches Elektronen-Synchrotron DESY, \cr
     $\quad \;$ Notkestrasse 85, 22607 Hamburg, Germany 
    }

\institution{HamMath}{$^{(1,3)}$ Fachbereich Mathematik,
 Universit\"at Hamburg, \cr
 $\quad \;$ Bundesstrasse\ 55,
 20146 Hamburg, Germany}

 \institution{Kings}{$^{(1,5)}$ Department of Mathematics,
 King’s College London,\cr
 $\quad \;\,$ Strand, London WC2R 2LS, United Kingdom}

\title{\textbf{Non-local charges from perturbed defects via SymTFT in 2d CFT}}

\authors{Federico Ambrosino
 \orcidlink{0000-0002-5617-5446}$^{(1,2)}$, 
Ingo Runkel \orcidlink{0000-0002-0496-3096}$^{(1,3)}$ and  G\'erard M.\ T.\ Watts \orcidlink{0000-0002-9066-2838}$^{(1,5)}$} 

\abstract{
We investigate non-local conserved charges in perturbed two-dimensional conformal field theories from the point of view of the 3d SymTFT of the unperturbed theory. In the SymTFT we state a simple commutation condition which results in a pair of compatible bulk and defect perturbations, such that the perturbed line defects are conserved in the perturbed CFT. In other words, the perturbed defects are rigidly translation invariant, and such defects form a monoidal category which extends the topological symmetries. 

As examples we study the A-type Virasoro minimal models $M(p,q)$. Our formalism provides one-parameter families of commuting non-local conserved charges for perturbations by a primary bulk field with Kac label $(1,2)$, $(1,3)$, or $(1,5)$, which are the standard integrable perturbations of minimal models. We find solutions to the commutation condition also for other bulk perturbations, such as $(1,7)$, and we contrast this with the existence of local conserved charges.
There has been recent interest in the possibility that in certain cases perturbations by fields such as $(1,7)$ can be integrable, and our construction provides a new  way in which integrability can be found without the need for local conserved charges.
}

    \maketitle
    \setcounter{tocdepth}{2}	

    \tableofcontents

    \newpage

\section{Introduction}

Topological defects are an important tool in understanding properties of quantum field theories.
We will be concerned with two dimensional theories and line defects in these theories. 
If one perturbs the QFT by some bulk field(s), a subset of topological line defects may persist along the deformation, revealing for example non-trivial information about the infrared fixed point (see e.g.\ \cite{Fredenhagen:2009tn,Gaiotto:2012np,Chang:2018iay,Komargodski:2020mxz,Bhardwaj:2023idu,Cordova:2024vsq,Nakayama:2024msv}).
In the present paper we will explore the idea that one should also consider non-topological defects and that these
contain interesting and non-trivial information about deformations of QFTs, and in particular may allow one to identify integrable perturbations. 
Our construction builds on the works \cite{Bazhanov:1994ft,Fioravanti:1995cq,Bazhanov:1996dr,Bazhanov:1996aq,Bazhanov:2001xm,Runkel:2007wd,Runkel:2010ym}.
Despite the name, there is no direct relation to early work on non-local charges in quantum field theories 
\cite{Luscher:1977uq,Zamolodchikov:1989rd,Bernard:1990ys}
which lie behind Yangian and affine quantum group symmetry of many models.

Below we give a quick summary of our construction using the 3d\,SymTFT of a 2d\,CFT and we present our findings in the example of Virasoro minimal models. 

\medskip

 We start from a 2d\,CFT $\cal C$ which we describe via the SymTFT picture (see
\cite{Kong:2020cie,
Gaiotto:2020iye,
Apruzzi:2021nmk, 
Freed:2022qnc,
Kaidi:2022cpf,
Bhardwaj:2023ayw,
Schafer-Nameki:2023jdn,
Bhardwaj:2023kri,
Carqueville:2023jhb}) 
as a 3d\,TFT with a conformal and a topological boundary condition. 
Concretely we will use the 3d\,TFT obtained by folding from the original TFT description of CFT correlators developed in \cite{Felder:1999mq,Fuchs:2002cm,Frohlich:2006ch}, see Section~\ref{sec:CFT-via-SymTFT}. The relation between these two points of view is briefly explained e.g.\ in \cite{Romaidis:2023zpx,Carqueville:2023jhb}.

The CFT $\cal C$ is perturbed by a relevant primary field $\varphi$, 
which in the SymTFT is represented by a line defect stretching between the conformal and the topological boundary of the 3d\,TFT, together with a choice of topological junction on the topological boundary,
\be
    \includegraphics[valign=c]{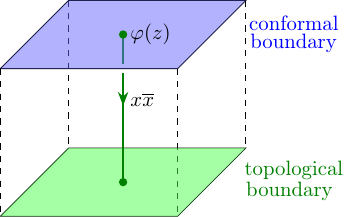}
\ee
The label of the line defect is the product $x\overline{x}$ of line defects $x$ and $\ov x$, where $x$ ends on a holomorphic field on the conformal boundary and $\overline x$ on an antiholomorphic field.\footnote{
In more detail, elementary line defects in this SymTFT are labelled by pairs $u \boxtimes v$, where $u = u \boxtimes \mathbf{1}$ labels a line that ends on a holomorphic field on the conformal boundary, and $v = \mathbf{1} \boxtimes v$ ends on an antiholomorphic field. Using this notation, the product of $u$ and $v$ is $uv = u \boxtimes v$, cf.\ \cite[Fig.\,3.4]{Carqueville:2023jhb}.
} 
We write $\cC(\mu\varphi)$ for $\cal C$ perturbed by $2i\mu\varphi$, $\mu \in \mathbb{C}$. The factor of $2i$ is a convention which reduces the factors of $i$ in our main application, alternatively we could have worked with $\mu' := 2 i \mu$ everywhere.

A topological line defect $\cD$ in the 2d\,CFT is represented by a line $\cD$ in the topological boundary. If the defect $\cD$ commutes with the perturbing field $\varphi$, it remains topological in $\cC(\mu\varphi)$ for all $\mu \in \mathbb{C}$. In the SymTFT, this condition takes the form
\be
\label{eq:intro-defect-transparent}
\includegraphics[valign=c]{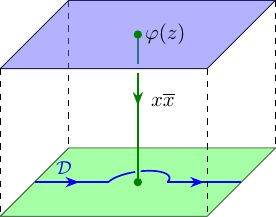} \qquad  \scalebox{1.5}{\textbf{=}}  \qquad \includegraphics[valign=c]{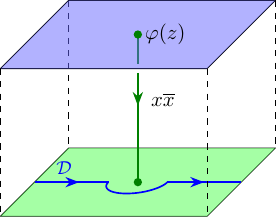}
\ee
Apart from topological defects, the CFT $\cal C$ contains another important class of defects, namely those which are invariant under rigid translations, i.e.\ commute with the Hamiltonian of $\cal C$. Such defects still allow for a non-singular fusion operation and have interesting algebraic properties \cite{Runkel:2007wd,Manolopoulos:2009np}. One can now ask if such translation invariant operators also exist in the perturbed theory $\cC(\mu\varphi)$.
To this end, denote by $\psi$ and $\ov\psi$ the holomorphic and antiholomorphic defect fields on the topological defect $\cD$ described in SymTFT as 
\be
\includegraphics[valign=c]{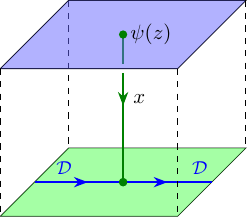} \qquad    \qquad \includegraphics[valign=c]{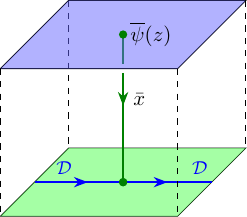}
\ee
Here, $\cD$ can be an elementary defect or a superposition of elementary defects, and $\psi$ can be a linear combination of fields that live on elementary defects and fields that interpolate them.

\begin{figure}
\begingroup
\addtolength{\jot}{1em}
\begin{equation*}
\begin{split}
    \includegraphics[valign=c]{Figs/Intro2a.pdf} \qquad  &\scalebox{2}{$-$}  \qquad \includegraphics[valign=c]{Figs/Intro2b.pdf} \\
     \scalebox{1.5}{\textbf{=}}\quad \includegraphics[valign=c]{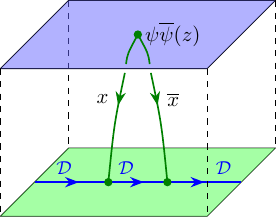} \qquad  &\scalebox{2}{$-$}  \qquad \includegraphics[valign=c]{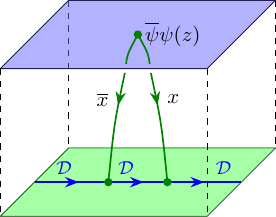} 
     \end{split}
\end{equation*}
\endgroup
    \caption{Bulk commutation condition in the SymTFT picture}
    \label{fig:intro-comm-con}
\end{figure}

A simple condition for the existence of translation invariant operators is the \textit{bulk commutation condition} \cite{Runkel:2010ym}, which in the SymTFT picture is given in Figure~\ref{fig:intro-comm-con}. Note that \eqref{eq:intro-defect-transparent} is a special case of this condition when one chooses $\psi=\ov\psi=0$.

If the bulk commutation condition is satisfied, we obtain a translation invariant operator in $C(\mu\varphi)$ as follows.
Denote by $\cD(\lambda\psi + \wt\lambda\ov\psi)$  the topological defect $\cD$ perturbed by the defect field $\lambda\psi + \wt\lambda\ov\psi$, 
then for $\lambda \wt\lambda = \mu$ the perturbed defect $\cD(\lambda\psi + \wt\lambda\ov\psi)$ commutes with the Hamiltonian of the perturbed theory $\cC(\mu \varphi)$ (see Section~\ref{sec:bulk-comm-cond}). We refer to $\cD(\lambda\psi + \wt\lambda\ov\psi)$ as a \textit{non-local conserved charge} -- ``non-local'' because it is obtained by integrating (anti)holomorphic fields which only exist on the defect $\cD$ and are not local currents in the bulk of the CFT.

Altogether, we obtain a one-parameter family  $\cD(\lambda\psi + \mu/\lambda\,\ov\psi)$ for $\lambda \in \mathbb{C}^\times$ of non-local conserved charges for $\cal C(\mu\varphi)$. In all the 
examples we consider in detail,  we can moreover show that the defect operators commute for different values of $\lambda$,
\begin{equation}\label{eq:commuting-defect-operators}
 \big[ \,  \cD(\lambda\psi + \mu/\lambda\ov\psi) \,,\,  \cD(\lambda'\psi + \mu/\lambda'\ov\psi) \, \big] \,=\, 0 \ .
\end{equation}
In other words, we obtain a one-parameter family of non-local conserved charges in involution in the perturbed CFT $\cal C(\mu\varphi)$, which is an indication that $\cal C(\mu\varphi)$ describes an integrable deformation of $\cal C$.

Such non-local conserved charges were first considered in the chiral free boson theory in \cite{Bazhanov:1994ft,Bazhanov:1996dr,Bazhanov:1996aq} and for chiral minimal models in \cite{Fioravanti:1995cq}. The formulation in terms of perturbed defects in full minimal models was given in \cite{Runkel:2007wd,Runkel:2010ym} using the TFT formulation of CFT correlators.

If we apply the above to the unperturbed theory, i.e.\ to $\cC(\mu \varphi)$ with $\mu=0$, we are lead to considering $\cD(\lambda\psi + \wt\lambda\ov\psi)$ with either $\lambda=0$ or $\wt\lambda=0$. Indeed, the defect $D$ perturbed by only $\psi$ or by only $\ov\psi$ commutes with the conformal Hamiltonian and provides families of non-local conserved charges in the CFT $\cal C$ \cite{Runkel:2007wd}.

\medskip

To summarise the discussion so far, 
in addition to the topological line defects $\mathcal{T}(\mu)_\mathrm{top}$ of $\mathcal{C}(\mu \varphi)$, one should consider all rigidly translation invariant defects $\mathcal{T}(\mu)$ in $\mathcal{C}(\mu \varphi)$. These share some of the good properties with $\mathcal{T}(\mu)_\mathrm{top}$, in particular their fusion is non-singular so that $\mathcal{T}(\mu)$ is a monoidal category (see Section~\ref{sec:rigid-trans}). The bulk commutation condition gives access to perturbatively defined  rigidly translation invariant defects $\mathcal{T}(\mu)_\mathrm{pert}$, and it turns out that these again close under fusion. We obtain inclusions of monoidal subcategories,
\begin{equation}
    \mathcal{T}(\mu)_\mathrm{top} ~\subset~
    \mathcal{T}(\mu)_\mathrm{pert} ~\subset~\mathcal{T}(\mu) \ .
\end{equation}
This contains the case $\mu=0$ of the unperturbed theory. We propose that considering $\mathcal{T}(\mu)$, rather than just $\mathcal{T}(\mu)_\mathrm{top}$, gives important additional information about the family of perturbed CFTs $C(\mu\varphi)$. For example, a rigidly translation invariant defect is expected to flow to a topological defect in the infrared theory as the requirement of conformal invariance combined with translation invariance is sufficient to prove a defect is topological, as we show in Section~\ref{sec:rigid-trans}.

In this paper we restrict our attention to two-dimensional QFTs, but rigidly translation invariant defects can be considered in $d$-dimensional QFTs as well. They would again have well-defined fusion and would form a $d$-categorical structure extending that of topological defects.

\medskip

Let us now turn to examples of the above construction. We focus on A-type minimal models $M(p,q)$. Here $p,q \ge 2$ are coprime integers.

Denote by $\mathcal{I}$ the set of Kac-labels for $M(p,q)$, that is all $(r,s)$ with $1 \le r < p$ and $1 \le s < q$ modulo the identification $(r,s) \sim (p-r,q-s)$. We write $h_{r,s}$ for the conformal weight of the primary state in the irreducible Virasoro representation corresponding to $(r,s)$. The primary bulk fields of $M(p,q)$ are labelled by $x = (r,s) \in \mathcal{I}$ and have conformal weights $(h_{r,s},h_{r,s})$. The elementary topological line defects are equally labelled by $a = (r,s) \in \mathcal{I}$ \cite{Petkova:2000ip, Fuchs:2002cm}, and the (anti)holomorphic defect fields which live on the topological defect $a$ are precisely those of weight $(h_y,0)$ or $(0,h_y)$ for $y$ occurring in the fusion $a \otimes a$, i.e.\ those for which $N_{aa}^{~y} \neq 0$.

In Table~\ref{tab:intro-solved} we list the cases of primary bulk fields $\varphi$ and topological defect $D$ that we investigate in some detail in Section~\ref{sec:charges-in-Vir}. Each produces a solution to the bulk commutation condition in Figure~\ref{fig:intro-comm-con} and -- at least in the cases where the perturbation does not require regularisation -- a one-parameter family of translation invariant defect operators $\cD(\lambda\psi + \mu/\lambda\,\ov\psi)$ in the minimal model $M(p,q)$ perturbed by $\mu\varphi$. We also verify that the corresponding defect operators commute for different value of $\lambda$ as in \eqref{eq:commuting-defect-operators}, indicating that the corresponding perturbations are integrable. We only list cases $(1,s)$, the cases $(s,1)$ can be obtained by exchanging $p \leftrightarrow q$.

The solutions include the perturbations by $(1,2)$, $(1,3)$ and $(1,5)$, which are the standard integrable deformations of minimal models. The perturbed defects do not require regularisation for $t<\frac43$, $t<\frac34$, and $t<\frac25$, respectively.
We note that the $(1,7)$-perturbations require regularisation in all cases listed in Table~\ref{tab:intro-solved}
(except $M(2,9)$). Still, the $(1,7)$-perturbation of the minimal model $M(3,10)$ has recently received considerable attention and is a new candidate for an integrable flow \cite{Klebanov:2022syt, Nakayama:2024msv,Katsevich:2024jgq,Ambrosino:2025xsv}, and it is interesting to observe that the bulk-commutation relation is satisfied in this case.

\begin{table}
\begin{center}
\renewcommand{\arraystretch}{1.4}
\begin{tabular}{cccc}
Minimal& perturbing& weights & topological defect
\\[-.4em]
model $M(p,q)$ & bulk field $\varphi$ & $h=\overline h$ of $\varphi$ & $\mathcal{D}$ in Figure~\ref{fig:intro-comm-con}
\\
\hline
$q \ge 3$ & $(1,2)$ & $h_{1,2} = \frac{3}{4} t - \frac12$ & $(1,1) \oplus (1,2)$
\\
$q \ge 4$ & $(1,3)$ & $h_{1,3} = 2t-1$ & $(1,2)$
\\
$q\ge 6$ & $(1,5)$ & $h_{1,5} = 6t-2$ & $(1,3)$\\
$q=9,10,18$ & $(1,7)$ & $h_{1,7} = 12t-3$ & $(1,5)$ \\
\end{tabular}
\end{center}

\caption{Examples of minimal models in which we solve the bulk commutation condition (Figure~\ref{fig:intro-comm-con}) and verify that the corresponding perturbed defects mutually commute. As is usual,  $t=p/q$. 
For the (1,7)-perturbation on the (1,5)-defect, the bulk commutation condition can only be solved for the given values of $q$. (But other solutions exist for different choices of topological defect.)}
\label{tab:intro-solved}
\end{table}

There are many more solutions to the bulk commutation condition than those listed in Table~\ref{tab:intro-solved}; we collect some of these in Section~\ref{ssec:furtherobs}.

\medskip

An indicator of integrability which has been extensively studied is the existence of higher spin conserved currents in the perturbed CFT (see e.g.\ \cite{Dorey:1996gd}).
A necessary condition for a holomorphic field $W(z)$ in the CFT to  survive as a conserved current when the CFT is perturbed by $\varphi(w)$ is that the first order pole in the OPE $W(z)\varphi(w)$ is a total derivative \cite{Zamolodchikov:1989hfa}. We analyse this condition in Section~\ref{ssec:localsearch}. The standard integrable perturbations $(1,2)$, $(1,3)$, $(1,5)$ are known to possess higher conserved charges related to exponents of affine Lie algebras. For the $(1,7)$-perturbation we find such higher conserved charges only in a few models, and in particular not in all the models that solve the bulk commutation condition. 

\bigskip

This paper is organised as follows. 
In Section~\ref{sec:SymTFT} we state the bulk commutation condition and give two criteria which imply that the corresponding perturbed defects mutually commute. We introduce the monoidal category of rigidly translation invariant defects.
In Section~\ref{sec:charges-in-Vir} we investigate the bulk commutation condition and mutual commutativity of the perturbed defects in perturbed Virasoro minimal models.
In Section~\ref{sec:local-charges} we systematically look for local conserved charges in the models studied in Section~\ref{sec:charges-in-Vir}. In the outlook 
in Section~\ref{sec:outlook} we collect some questions left unanswered in this work and further future research directions.

\section{SymTFT and non-local conserved charges}\label{sec:SymTFT}

To understand properties of a quantum field theory in $d$ dimensions, such as its locality properties and topological symmetries, it is useful to represent the theory as a $(d+1)$-dimensional topological field theory together with a non-topological boundary and a topological boundary, see e.g.\ \cite{Kong:2020cie,
Gaiotto:2020iye,
Apruzzi:2021nmk,
Freed:2022qnc,
Kaidi:2022cpf,
Bhardwaj:2023ayw,
Schafer-Nameki:2023jdn,
Bhardwaj:2023kri,
Carqueville:2023jhb}). We will apply this picture in the specific situation that the QFT in question is a two-dimensional conformal field theory, so that the SymTFT is three-dimensional.

Below we will first describe bulk and defect fields in the SymTFT picture, in particular making the relation to the original ``unfolded'' or ``chiral'' TFT description of CFT correlators in \cite{Felder:1999mq,Fuchs:2002cm,Frohlich:2006ch}. Then we formulate the bulk commutation condition from \cite{Runkel:2010ym} in SymTFT and chiral TFT language. We explain how it leads to non-local conserved charges in the perturbed theory and state two criteria which ensure that the defect operators mutually commute. Finally, we briefly discuss the extension from topological line defects to rigidly translation invariant defects.

\subsection{CFT with defects via SymTFT}\label{sec:CFT-via-SymTFT}

We present the fields and defects that we will we need in three ways: on the world sheet, in SymTFT, and in chiral TFT where holomorphic and antiholomorphic fields are on separate boundaries of the 3d\,TFT.

\begin{figure}
    \centering
    \includegraphics[width = \textwidth]{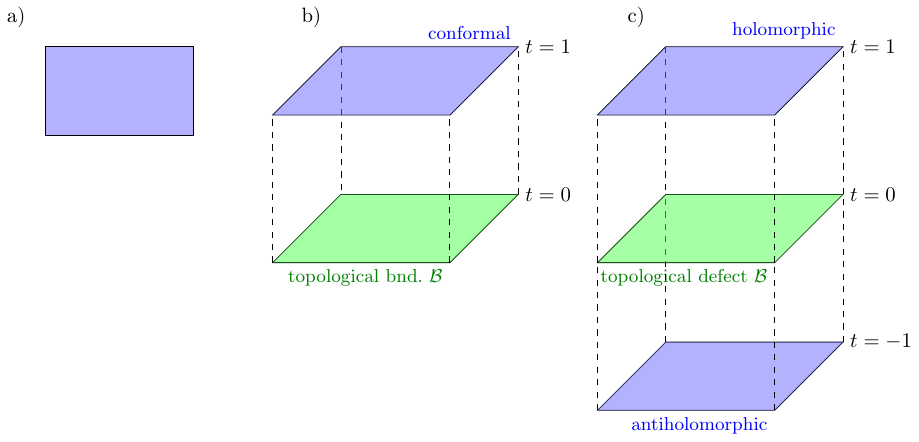}
    \caption{a) A patch of the surface $\Sigma$ on which the 2d\,CFT is defined.
    b) The SymTFT presentation on $\Sigma \times [0,1]$ with topological boundary condition $\mathcal{B}$.
    c) The chiral TFT presentation on $\Sigma \times [-1,1]$ with a topological surface defect with defect condition $\mathcal{B}$ at $\Sigma \times \{0\}$.}
    \label{fig:geometry}
\end{figure}

\medskip

Let us start with the \textbf{geometric setup}, cf.\ Figure~\ref{fig:geometry}. Let $\Sigma$ be a surface with complex structure on which we can evaluate correlators of the 2d\,CFT.
Then in SymTFT, this becomes $\Sigma \times [0,1]$, where $\Sigma \times \{1\}$ is a copy of $\Sigma$ with its complex structure, and $\Sigma \times \{0\}$ is a topological boundary and carries a boundary condition $\mathcal{B}$. Different choices of $\mathcal{B}$ give different choices of full (or absolute)  2d\,CFT upon collapsing the interval $[0,1]$. In the chiral TFT description, we consider $\Sigma \times [-1,1]$. The boundaries $\Sigma \times \{1\}$ and $\Sigma \times \{-1\}$ are both non-topological and carry a complex structure, with the former supporting the holomorphic and the latter the antiholomorphic local degrees of freedom. The holomorphic and antiholomorphic boundaries are separated by a topological surface defect placed at $\Sigma \times \{0\}$ and which we also label by $\mathcal{B}$ for a reason that will become clear momentarily. The interpretation of the construction in \cite{Fuchs:2002cm} in terms of surface defects is due to \cite{Kapustin:2010if,Fuchs:2012dt}. For more on surface defects in these types of 3d\,TFTs see \cite{Carqueville:2017ono,Koppen:2021kry}.

To obtain the SymTFT picture from the chiral TFT picture, we fold the chiral TFT picture at $\Sigma \times \{0\}$, i.e.\ we identify $t \sim -t$ in the interval $[-1,1]$. This creates a topological boundary at the fixed point $t=0$ and a non-topological boundary with both holomorphic and antiholomorphic fields at $t=1$. The topological boundary condition at $t=0$ is obtained from the topological defect condition $\mathcal{B}$.

\medskip

\begin{figure}[t]
   \centering
\begin{equation*}
\begin{split}
    \includegraphics[valign=c]{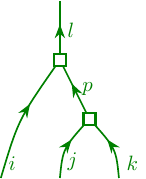}   =  \sum_q \F{ijk}{l}{pq} \includegraphics[valign=c]{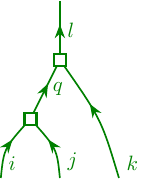}  \qquad  \includegraphics[valign=c]{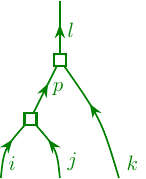}   = \sum_q \G{ijk}{l}{pq} \includegraphics[valign=c]{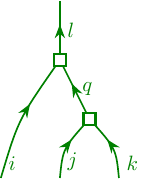} \\
     \includegraphics[valign = c]{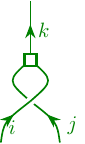} = \R{ij}{k} \includegraphics[valign = c]{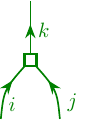}\qquad\qquad  \includegraphics[valign = c]{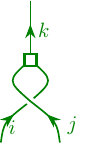} = \frac{1}{\R{ji}{k}} \includegraphics[valign = c]{Figs/R0.pdf}\qquad\qquad \includegraphics[valign = c]{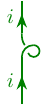} = \theta_i \; \; \includegraphics[valign = c]{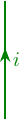}\\
     \includegraphics[valign = c]{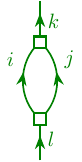} = \delta_{kl} \, {\rm N}_{ij}^k \;\; \includegraphics[valign=c]{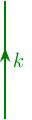}\qquad\qquad  \includegraphics[valign=c]{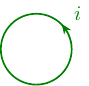} = \includegraphics[valign=c]{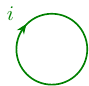} = {\rm dim} (i)\qquad \qquad
     \end{split}
\end{equation*}
    \caption{Local identities for line defects in the chiral TFT for the modular fusion category $\mathcal{F}$. Here, $i,j,k,\dots \in \mathrm{Irr}(\mathcal{F})$ denote simple objects, and the small boxes denote the basis element $\lambda_{ij}^k : i\otimes j \to k$ and its dual $k \to i\otimes j$.
    }
    \label{fig:TFT-line-rules}
\end{figure}

For our explicit computations, the chiral TFT picture will be the most convenient. Technically, the chiral TFT is obtained as the Reshetikhin-Turaev TFT for the modular fusion category $\mathcal{F} := \mathrm{Rep}\mathcal{V}$, the category of representations of the chiral algebra (a rational vertex operator algebra) $\mathcal{V}$.
Line defects in the chiral TFT are labelled by objects of $\mathcal{F}$, i.e.\ by representations of the chiral algebra. Write $\mathrm{Irr}(\mathcal{F})$ for a choice of simple objects (irreducible representations) in $\mathcal{F}$. We denote by $\mathbf{1} \in \mathrm{Irr}(\mathcal{F})$ the unit object (vacuum representation). For $i,j,k \in \mathrm{Irr}(\mathcal{F})$, the fusion coefficients $N_{ij}^{~k}$ give the multiplicity of $k$ in the product $i \otimes j$. For simplicity we will make the
\begin{quote}
\textbf{Assumption:} $N_{ij}^{~k} \in \{0,1\}$ for all $i,j,k \in \mathrm{Irr}(\mathcal{F})$.
\end{quote}
This avoids multiplicity labels and is satisfied in Virasoro minimal models, which are the explicit examples we consider later on.
If $N_{ij}^{~k}=1$, we fix once and for all a non-zero vector $\lambda_{ij}^k \in \mathrm{Hom}_{\mathcal{F}}(i \otimes j,k)$, which is hence a basis of that one-dimensional space. The local rules to manipulate networks of line defects in the chiral TFT are collected in Figure~\ref{fig:TFT-line-rules} -- these agree with the conventions used in \cite{Fuchs:2002cm}.

\medskip

For the rest of this paper, on the side of the chiral TFT we will only consider the trivial surface defect, as this is the situation we will use later on. In SymTFT we keep the picture general.

\medskip

\begin{figure}
    \centering
    \includegraphics[width = \textwidth]{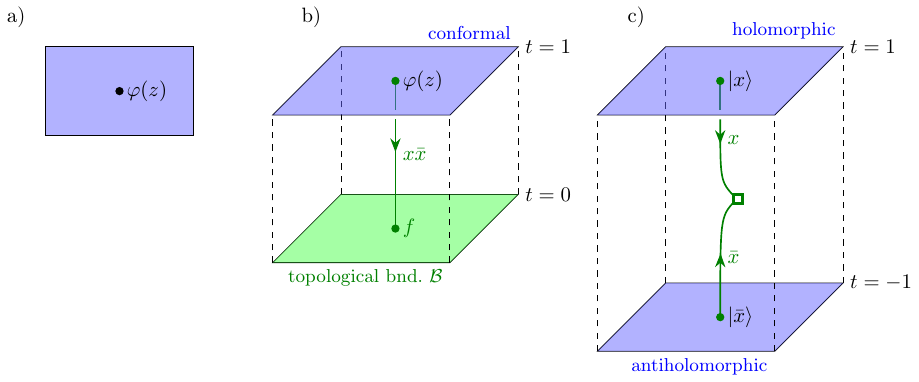}
    \caption{a) CFT: A patch of the surface $\Sigma$ with an insertion of the bulk field $\varphi(z)$ of the 2d\,CFT.
    b) SymTFT presentation. c) Chiral TFT presentation.
    Note that we are now assuming the surface defect in Figure~\ref{fig:geometry}\,c) to be trivial. }
    \label{fig:bulk-field}
\end{figure}

Let us now describe \textbf{bulk fields} $\varphi(z)$ in the 2d\,CFT,\footnote{
We take the notation $\varphi(z)$ to mean the field $\varphi$ inserted at point $z$, and that there is no implicit statement about $\varphi(z)$ being holomorphic or not. We will mostly avoid the notation $\varphi(z,\ov z)$ which is often used in 2d\,CFT in this context. 
}
cf.\ Figure~\ref{fig:bulk-field}.
On the SymTFT side, the bulk field is described by a non-topological point insertion $\varphi$ on the conformal boundary and a topological point insertion $f$ on the topological boundary.
In the chiral TFT, the bulk field consists of two line defects pointing away from the boundary which meet at a topological junction which we label by our standard basis $\lambda_{x\overline x}^{\mathbf{1}}$, and which we represent by a small box.
The endpoints of the line defect on the holomorphic and antiholomorphic boundary and are labelled by states (primary in our application) $|x\rangle \in x$ and $|\overline x\rangle \in \overline x$, so that $\varphi = |x\rangle \otimes_{\mathbb{C}} |\overline x\rangle$ with conformal weights $(h,\overline h) = (h_x,h_{\overline x})$.

\begin{figure}
    \centering
    \includegraphics[width = \textwidth]{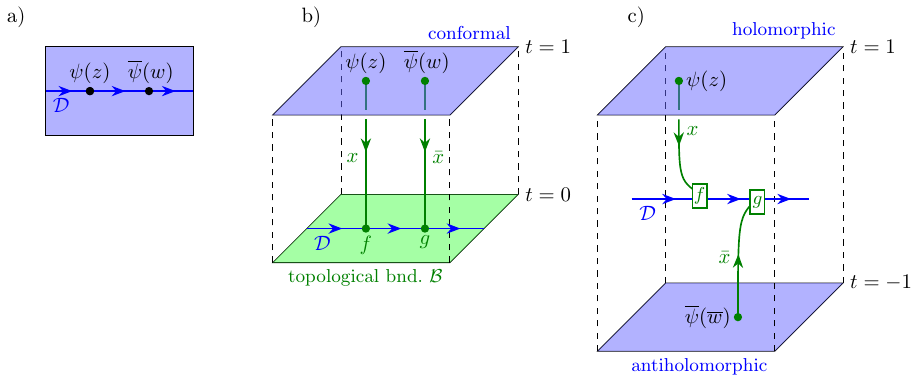}
    \caption{a) CFT: A patch of the surface $\Sigma$ with an insertion of a topological line defect $D$, a holomorphic defect field $\psi$ and an antiholomorphic defect field $\ov\psi$.
    b) SymTFT presentation. Here we assume that the endpoints of $x$ and $\overline x$ on the conformal boundary depend holomorphically (resp.\ antiholomorphically) on the insertion point. c) Chiral TFT presentation. 
    }
    \label{fig:defect-field}
\end{figure}

\medskip

\textbf{Topological line defects} in the 2d\,CFT are given by line defects $\cD$ on the topological boundary of the SymTFT, and simply by line defects $\cD$ in the chiral TFT (as we assume the surface defect to be trivial). We will only need holomorphic and antiholomorphic \textbf{defect fields}, not general defect fields. Their description is as for bulk fields, but now the TFT line defects which are along the interval direction need to end on the line defect, see Figure~\ref{fig:defect-field}. On the side of the chiral TFT, the defect is labelled by a not necessarily simple object $\cD \in \mathcal{F}$, and $f : x \otimes \cD \to \cD$, $g : \cD \otimes \overline x \to \cD$ are morphisms in $\mathcal{F}$. The chiral TFT has the advantage of clearly showing which fields are holomorphic, and which fields are antiholomorphic (and which are neither, as the bulk field shown in Figure~\ref{fig:bulk-field}\,c).

\subsection{The bulk commutation condition}\label{sec:bulk-comm-cond}

Consider the perturbation of the CFT $\mathcal{C}$ by a a spin-less bulk field $\varphi$, that is, a bulk field with conformal dimensions $h = \overline h$. We take $\varphi$ to transform in the representations $x$ and $\overline x$ of the holomorphic and antiholomorphic copy of the chiral algebra $\mathcal{V}$, respectively. Its representation in SymTFT and chiral TFT is as in Figure~\ref{fig:bulk-field}.

Denote by $\mathcal{C}(\mu \varphi)$ the CFT $\mathcal{C}$ perturbed by $2i\mu \varphi$, $\mu \in \mathbb{C}$ (the normalisation factor of $2i$ will be convenient later). This includes the case $\mu=0$, which is just the unperturbed theory. We consider $\mathcal{C}(\mu \varphi)$ on cylinder of circumference $L$. The Hamiltonian of $\mathcal{C}(\mu \varphi)$ is
\begin{equation}
	H(\mu) = H_0 + H_\mathrm{pert}(\mu)
	~~,\quad
	H_0 = \frac{2\pi}{L} \Big( L_0 + \ov L_0 - \frac{c}{12} \Big)
	~~,\quad
	H_\mathrm{pert}(\mu) = 2i \mu \int_0^L \hspace{-.4em} \varphi(s) \, ds  \ .
\end{equation}

\begin{figure}
	\centering
	\includegraphics[width = 0.7\textwidth]{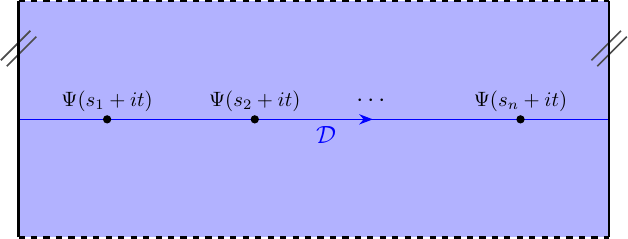}
	\caption{A contribution to the perturbative expansion of $\cD(\lambda \psi + \wt\lambda \ov\psi)$ placed around a cylinder of circumference $L$. Here $\Psi = \lambda \psi + \wt\lambda \ov\psi$ and the positions $0< s_1 < \dots < s_n<L$ of the fields are being integrated over. 
    }
	\label{fig:perturbing term}
\end{figure}

Next we choose a topological defect $\cD$ of the unperturbed CFT $\mathcal{C}$, as well as two defect fields $\psi$ and $\ov\psi$, where $\psi$ is holomorphic and transforms in representation $x$, while $\ov\psi$ is antiholomorphic and transforms in representation $\ov x$. Their TFT representation is as in Figure~\ref{fig:defect-field}. We write $\cD(\lambda \psi + \wt\lambda \ov\psi)$ for the defect $\cD$ perturbed by the field $\Psi = \lambda \psi + \wt\lambda \ov\psi$. Explicitly,
\begin{equation}
	\cD(\lambda \psi + \wt\lambda \ov\psi;t)
	= \exp\!\Big( \int_0^L \big(\lambda \psi(s+it) + \wt\lambda \ov\psi(s+it)\big) \, ds \Big) \ ,
\end{equation}
where we include the position $t$ of the defect loop along the cylinder, and $s$ is the position on the defect. See Figure~\ref{fig:perturbing term} for one term in the perturbative expansion.
If we would like the integrals to converge without further regularisation, we should impose 
\be\label{eq:convergence-necessary-cond}
    2h - h_\mathrm{min} \,<\,1 \ ,
\ee
where $h_\mathrm{min}$ is the smallest weight of primary fields that occur on the OPE $\psi(x)\psi(y) = (x-y)^{h_\mathrm{min}-2ph} \phi(y) + \dots$ (and so in particular has to be a holomorphic defect field on $\mathcal{D}$). As the CFT can be non-unitary, $h_\mathrm{min}<0$ is possible.

This is a stronger condition than that of $\psi$ being relevant, which would be $h<1$. Note also that more stringent constraints than \eqref{eq:convergence-necessary-cond} can arise by considering iterated OPEs.

\medskip

\begin{figure}[t]
\begingroup
\addtolength{\jot}{1em}
\centering
\begin{equation*}
\begin{split}
    &\includegraphics[scale=1,valign=c]{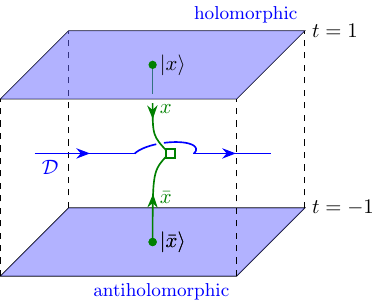}   
    \scalebox{2}{$-$}  \qquad \includegraphics[scale=1,valign=c]{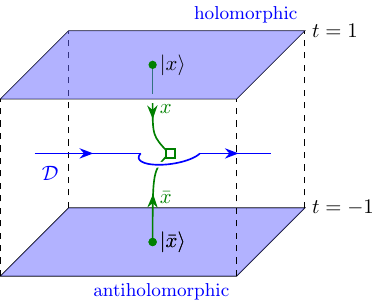} \\[-.5em]
     & \scalebox{1.5}{\textbf{=}}\qquad \includegraphics[scale=1,valign=c]{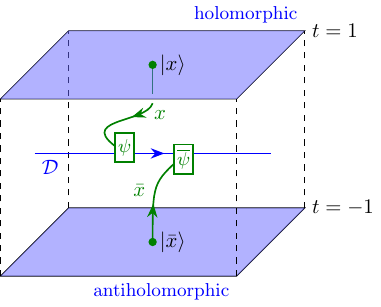}  
     \scalebox{2}{$-$}  \qquad \includegraphics[scale=1,valign=c]{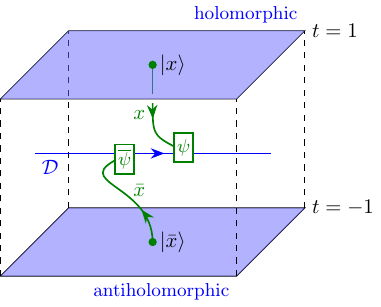} 
     \end{split}
\end{equation*}
\endgroup
	\caption{The bulk commutation condition expressed in terms of chiral TFT with trivial surface defect -- the SymTFT formulation was already given in Figure~\ref{fig:intro-comm-con}. 
    } 
	\label{fig:commutation_condition_chiral}
\end{figure}

We would like $\cD(\lambda \psi + \wt\lambda \ov\psi)$ to be rigidly translation invariant in the perturbed CFT $C(\mu\varphi)$. In other words, we would like it to commute with the perturbed Hamiltonian:
\begin{equation}\label{eq:def-comm-ham}
	\big[ \,H(\mu) \,,\, \cD(\lambda \psi + \wt\lambda \ov\psi) \, \big] \,=\, 0 \ .
\end{equation}
There is a simple condition which implies \eqref{eq:def-comm-ham} order by order in perturbation theory, and which is formulated in the CFT $\mathcal{C}$, not in the perturbed theory:
\begin{align}
\mu \,[ \cD , \varphi(p) ] \,=\, \lambda \wt\lambda \, [ \psi(p) , \ov\psi(p) ]\;. 
\label{eq:full-bulk-comm}
\end{align}
Here, the first commutator is understood as being taken along the cylinder, and the second along the defect; the commutators are taken in the CFT, not in the perturbed theory, and $\mathcal{D}$ is the unperturbed topological defect.

If $\mu=0$, this says that chiral defects (i.e.\ only one of $\lambda$ and $\wt\lambda$ is non-zero) are translationally invariant in the undeformed CFT \cite{Runkel:2007wd} (see also \cite{Bazhanov:1994ft,Konik:1997gx,Constantin_Bachas_2004})
as are defects for which the defect fields commute
(as happens for example in the fusion product of deformed defects). 
If neither of the commutators in \eqref{eq:full-bulk-comm} vanishes, then we can rescale the fields such that \eqref{eq:full-bulk-comm} is equivalent to the two conditions
\begin{align}
    \label{eq:comm-conn-commutator}
	\big[ \cD , \varphi(p) \big] &= \big[ \psi(p) , \ov\psi(p) \big] \;,
    \\
    \lambda \wt\lambda & = \mu
    \;.\label{eq:lamlammu}
\end{align}
We call the condition \eqref{eq:comm-conn-commutator} the \textit{bulk commutation condition}.
In terms of SymTFT, the bulk commutation condition 
is given in Figure~\ref{fig:intro-comm-con}. 
The chiral TFT formulation is shown in Figure~\ref{fig:commutation_condition_chiral} (with the surface defect in Figure~\ref{fig:geometry} taken to be trivial).

It is shown in \cite{Runkel:2010ym} (and in more generality in \cite[App.\,A.2]{Buecher:2012ma}) that
\eqref{eq:comm-conn-commutator} and 
\eqref{eq:lamlammu} imply \eqref{eq:def-comm-ham}.
We will briefly review the argument below.
Implicit in this argument is the assumption that the integrals in $\cD(\lambda \psi + \wt\lambda \ov\psi)$ are well-defined at each order in $\lambda,\wt\lambda$, cf.\ \eqref{eq:convergence-necessary-cond}. Otherwise one would need a regularisation procedure which is compatible between bulk and defect and we will not investigate this here. 

The case $\mu=0$ was already discussed above, so we now take $\mu \neq 0$ and restate 
\eqref{eq:comm-conn-commutator} in this case:
\begin{equation}\label{eq:comm-con-mu-nonzero}
\big[ \cD , \varphi(p) \big] \,=\, \big[ \psi(p) , \ov\psi(p) \big]
\quad \Rightarrow \quad
\big[ H(\mu) , \cD(\lambda \psi + \mu/\lambda  \ov\psi) \big] \,=\, 0 \ .
\end{equation}
Since $\lambda \in \mathbb{C}^\times$ is still arbitrary, the bulk commutation condition always provides a one-parameter family of rigidly translation invariant defects in $\mathcal{C}(\mu \varphi)$. We will address the question if the defect operators mutually commute for different values of $\lambda$ in Section~\ref{sec:def-op-commute}.

\medskip

Let us now quickly recap the argument leading to \eqref{eq:comm-con-mu-nonzero} as given in \cite{Runkel:2010ym,Buecher:2012ma}. Define $\Theta = \lambda \psi - \wt\lambda \ov\psi$ and $\Psi = \lambda \psi + \wt\lambda \ov\psi$ (as in Figure \ref{fig:perturbing term}). Observe that
\be\begin{aligned}\label{eq:Psi-to-Theta}
	\frac{\partial}{\partial t} \Psi(s+it)
	&=
	\frac{\partial}{\partial t} \Big( \psi(s+it) + \ov\psi(s+it) \Big)
	\\
	& =
	- i
	\frac{\partial}{\partial s} \Big( \psi(s+it) - \ov\psi(s+it) \Big)
	= - i
\frac{\partial}{\partial s} \Theta(s+it)
	\ ,
\end{aligned}\ee
because $\psi$ is holomorphic and $\ov\psi$ antiholomorphic.
While the OPE of $\Psi$ and $\Theta$ may be singular, for the difference of the two orders to take the OPE in we have
\begin{equation}\label{eq:PsiTheta-nonsing}
	\Theta(p+\varepsilon)\Psi(p) - \Psi(p+\varepsilon)\Theta(p)
	~\xrightarrow{~\varepsilon \to 0~}~
	2 \lambda \wt\lambda \big(\, \psi(p)\ov\psi(p) - \ov\psi(p)\psi(p) \, \big) \ .
\end{equation}
Using this, we compute
{\allowdisplaybreaks
\begin{align}\label{eq:comm-conn-derivation}
&	\Big[ H_0 \,,\,
	\int_{0 \le s_1 < \cdots < s_n \le L} 
    \Psi(s_1) \cdots \Psi(s_n)
	\,d^ns
	\Big]
	\nonumber	\\
&\overset{(1)}=
	\frac{\partial}{\partial t}
\int_{0 \le s_1 < \cdots < s_n \le L} \Psi(s_1+it) \cdots \Psi(s_n+it)
	\,d^ns
\Big|_{t=0}
	\nonumber	\\
&\overset{(2)}=
	-i\sum_{k=1}^n \int_{0 \le s_1 < \cdots < s_n \le L} 
    \Psi(s_1)\cdots  \frac{\partial}{\partial s_k} \Theta(s_k) \cdots \Psi(s_n)
	\,d^ns
	\nonumber	\\
&\overset{(3)}=
-i\sum_{k=1}^{n-1} \int_{0 \le s_1 < \cdots < s_{n-1} \le L} 
    \hspace{-2em}
\Psi(s_1)\cdots  \big(-\Psi(s_k) \Theta(s_k) + \Theta(s_k) \Psi(s_k)  \big)\cdots \Psi(s_{n-1})
	\,d^{n-1}s
	\nonumber	\\
&\overset{(4)}=
-2i \lambda \wt\lambda \sum_{k=1}^{n-1} \int_{0 \le s_1 < \cdots < s_{n-1} \le L} 
    \hspace{-2em}
\Psi(s_1)\cdots  \big(\psi(s_k)\ov\psi(s_k) - \ov\psi(s_k)\psi(s_k) \big) \cdots \Psi(s_{n-1})
	\,d^{n-1}s
	\nonumber	\\
&\overset{(5)}=
2i \lambda \wt\lambda \sum_{k=1}^{n-1} \int_{0 \le s_1 < \cdots < s_{n-1} \le L} 
\big[ \,\varphi(s_k) \,,\, \Psi(s_1)\cdots \widehat{\Psi(s_k)} \cdots \Psi(s_{n-1}) \, \big]
	\,d^{n-1}s
	\nonumber	\\
&=
2i \lambda \wt\lambda \Big[ \int_0^L \hspace{-0.3em}\varphi(r)\,dr \,,\, \int_{0 \le s_1 < \cdots < s_{n-2} \le L} \Psi(s_1)\cdots  \Psi(s_{n-2})
	\,d^{n-2}s
\,\Big] \ .
\end{align}}Step (1) follows as all expressions above are in the unperturbed theory $\mathcal{C}$ and $H_0$ generates translations along the cylinder. Step (2) is \eqref{eq:Psi-to-Theta}. In step (3) the integration over $s_k$ is carried out and we reorganise the sum to arrive at the non-singular combination \eqref{eq:PsiTheta-nonsing}. Here one should really work with a cut-off regularisation and take the cut-off to zero at the end. This has been done in \cite[App\,A.2]{Buecher:2012ma} to which we refer for details.
Step (4) is  \eqref{eq:PsiTheta-nonsing} and step (5) finally makes use of the bulk commutation condition \eqref{eq:comm-conn-commutator}. The hat indicates that $\Psi(s_k)$ is omitted.

Altogether, this establishes \eqref{eq:comm-con-mu-nonzero} order by order in $\mu = \lambda \wt\lambda$.

\subsection{The bulk commutation condition in a basis}

Here we give an explicit expression for the bulk commutation condition using the chiral TFT presentation in Figure~\ref{fig:commutation_condition_chiral} and the relations listed in Figure~\ref{fig:TFT-line-rules}. Let
\begin{equation}
	\cD = \bigoplus_{a \in \cD} a
\end{equation}
be an object in $\mathcal{F}$, with $a \in \mathrm{Irr}(\mathcal{F})$.  Recall that via the chiral TFT description in Figure~\ref{fig:defect-field}, $\cD$ represents a topological defect of the CFT $\mathcal{C}$ which is a superposition of the elementary defects $a \in \cD$.

We expand the holomorphic and antiholomorphic defect fields in representation $x$ and $\overline x$ in a basis as
\be\label{eq:psi-barpsi-expansion}
\psi = \sum_{a,b \, \in \,\cD} \kappa_{ab} \; \; \raisebox{0.42 cm}{\includegraphics[valign = c]{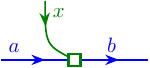}}\; \comma \qquad \ov\psi = \sum_{a,b \, \in \,\cD} \wt\kappa_{ab} \; \; \raisebox{-0.3 cm}{\includegraphics[valign = c]{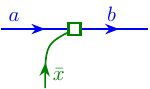}}\; \comma 
\ee
with $\kappa_{ab}, \wt\kappa_{ab} \in \mathbb{C}$.
The bulk field $\varphi$ is represented as in Figure~\ref{fig:bulk-field}\,c).
The bulk commutation condition in Figure \ref{fig:commutation_condition_chiral} reads
\be
\includegraphics[valign = c]{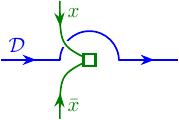} -
 \includegraphics[valign = c]{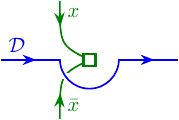} ~=~ 
 \includegraphics[valign = c]{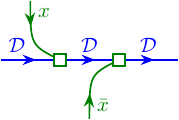}-
 \includegraphics[valign = c]{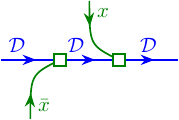} \ .
 \label{eq:16}
\ee
Expanding this into its simple components gives that for all $a,c \in \cD$,
\begin{align}
	&\delta_{a,c}
	 \left[\includegraphics[valign = c]{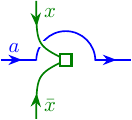} -
	\includegraphics[valign = c]{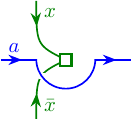} \right] = \sum_{b\in\cD} \kappa_{ab}\wt\kappa_{bc} \;\includegraphics[valign = c]{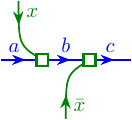}
	-\sum_{d\in\cD} \wt\kappa_{ad}\kappa_{dc} \;
	\includegraphics[valign = c]{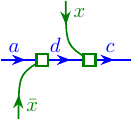}
\nonumber\\
	\label{eq:commutator}
	&\qquad \qquad \qquad\qquad= \sum_{b \in {\rm Irr}(\mathcal{F})} \left[ \delta_{b\in \cD} \;\kappa_{ab}\wt\kappa_{bc} - \sum_{d\in \cD} \F{xa \overline x}{c}{db} \, \wt\kappa_{ad} \kappa_{dc} \right] \; \includegraphics[valign =c]{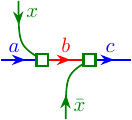}
    \end{align}
where the last equality follows from applying an F-move (Figure~\ref{fig:TFT-line-rules}).
Composing \eqref{eq:commutator}  with
\be\label{eq:bulk-comm-compose-with}
\includegraphics[valign = c]{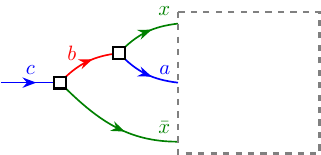}
\ee 
for $b \in {\rm Irr}(\mathcal{F})$
one obtains the following equivalent formulation of the bulk commutation condition: For all $a,c \in \cD$ and all $b \in {\rm Irr}(\mathcal{F})$ with $b \in a \otimes x$ and $b \in c \otimes x$ (so that the diagram in \eqref{eq:bulk-comm-compose-with} is non-zero) we must have
\be\label{eq:commutation-condition-basis}
\boxed{
	\delta_{ac}\,\left( \R{xa}{b} - \frac{1}{\R{ax}{b}}  \right) \F{ax\overline x}{a}{1b} \,=\, \delta_{b\in\cD}\, \kappa_{ab}\wt\kappa_{bc} - \sum_{d\in\cD} \F{xa \overline x}{c}{db} \, \wt\kappa_{ad} \kappa_{dc} } 
\ee
This is the most important equation of this section, and we will study some of its solutions in the case of minimal models in Section~\ref{sec:charges-in-Vir}.

\subsection{Fusion of defects and commutativity}\label{sec:def-op-commute}

Consider a topological defect $\cD = \bigoplus_{a \in \cD} a$ in $\mathcal{C}$ and suppose we can find a solution to the bulk commutation condition for $\cD$ for a given bulk perturbation $\mathcal{C}(\mu\varphi)$. In this section we give two (sufficient but not necessary) criteria for the one-parameter family of perturbed defects to be mutually commuting. 
More generally, one can consider two different perturbed defects, say $\cD(\lambda \psi + \mu/\lambda  \ov\psi)$ and $\cD'(\lambda' \psi' + \mu/\lambda'  \ov\psi{}')$. The parameter $\mu$ has to stay the same so that both defects are rigidly translation invariant in the same theory $\mathcal{C}(\mu \varphi)$. 
We then ask if, for $\lambda,\lambda' \in \mathbb{C}^\times$,
\begin{equation}\label{eq:do-defects-commute}
    \big[\, \cD(\lambda \psi + \mu/\lambda \ov\psi) \,,\, \cD'(\lambda' \psi' + \mu/\lambda'  \ov\psi{}') \,\big] ~=~ 0 \ .
\end{equation}
Both criteria we discuss use the fusion of the two perturbed defects. This is given by taking the fusion $\cD \otimes \cD'$ of the corresponding topological defects and perturbing by
\begin{equation}\label{eq:pert-field-on-fusion}
\lambda \psi + \lambda' \psi' + \mu/\lambda  \ov\psi + \mu/\lambda'  \ov\psi{}' \ ,    
\end{equation}
where it is understood that $\psi, \ov\psi$ are inserted on $\cD$, and $\psi'$, $\ov\psi{}'$ on $\cD'$.

\begin{figure}
	\begin{equation*}
\begin{split}
    &\includegraphics[valign=c]{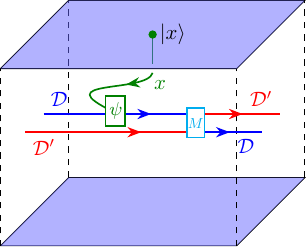}    \qquad 
 \scalebox{2}{$+$}  \qquad \includegraphics[valign=c]{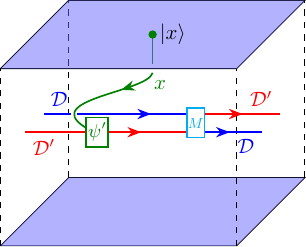} \\
&\scalebox{2}{$=$}  \qquad  \includegraphics[valign=c]{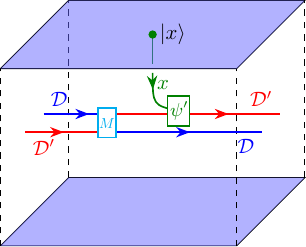}    \qquad 
 \scalebox{2}{$+$}  \qquad \includegraphics[valign=c]{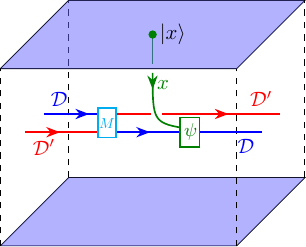} \quad  \\  \\ \\ \\
 &    \includegraphics[valign=c]{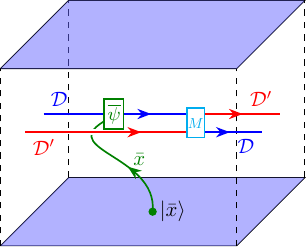}    \qquad 
 \scalebox{2}{$+$}  \qquad \includegraphics[valign=c]{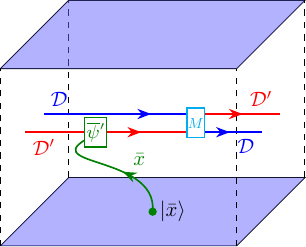} \\
&\scalebox{2}{$=$}  \qquad  \includegraphics[valign=c]{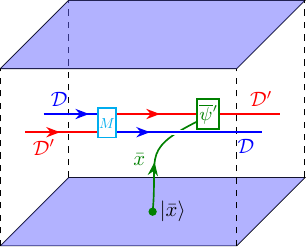}    \qquad 
 \scalebox{2}{$+$}  \qquad \includegraphics[valign=c]{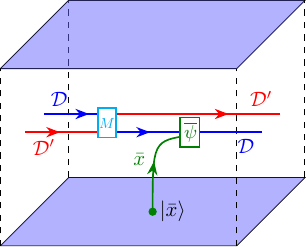}  \\
     \end{split}
\end{equation*}
    \caption{The two conditions for a topological junction $M  : \cD \otimes \cD' \to \cD' \otimes \cD$ to commute with the defect perturbation.
}
	\label{fig:train-track-condition}
\end{figure}

\subsubsection{Commuting defects via a topological junction}\label{sec:commute-via-M-matrix}

The first criterion is a version of the standard construction used in integrable lattice models to commute two transfer matrices past each other. In the present setting we are looking for a topological junction  $M = M(\lambda,\lambda') : \cD \otimes \cD' \to \cD' \otimes \cD$ which commutes with \eqref{eq:pert-field-on-fusion}. This leads to two conditions, one for the holomorphic fields and one for the antiholomorphic fields, whose chiral TFT presentation is shown in Figure~\ref{fig:train-track-condition}. If they are satisfied, the junction $M(\lambda,\lambda')$ remains topological in the perturbed theory. If the junction $M(\lambda,\lambda')$ has an inverse, one can insert a pair $\mathrm{id} = M M^{-1}$ next to each other and move $M$ around the defect circle to arrive at $M^{-1} M = \mathrm{id}$ and with the two defects $\cD(\lambda \psi + \mu/\lambda  \ov\psi)$ and $\cD'(\lambda' \psi' + \mu/\lambda'  \ov\psi{}')$ exchanged.

In summary, the first criterion states that if we can find an $M(\lambda,\lambda')$ which commutes with the perturbation as in Figure~\ref{fig:train-track-condition} and which is invertible, then \eqref{eq:do-defects-commute} holds.

\medskip

For explicit computations it is useful to express the above condition in a basis. For simplicity we assume that the elementary defects contained in $\cD$ and $\cD'$ have no multiplicities. Then the most general ansatz for $M$ is 
\begin{equation}\label{eq:ansatz-M}
    \includegraphics[valign=c]{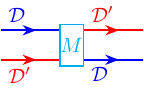} = \sum_{\substack{a,d\, \in \cD \\ b',c'\, \in \,\cD'\\ m \in \, {\rm Irr} F }} M_{ab'}^{c'd}(m) \;\; \includegraphics[valign = c]{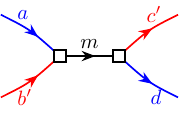}
    \quad .
\end{equation}
Next substitute this ansatz into the first condition in Figure~\ref{fig:train-track-condition}, together with the expansions of $\psi$ and $\ov\psi$ in \eqref{eq:psi-barpsi-expansion}, and compose with 
\begin{equation}
    \includegraphics[valign = c]{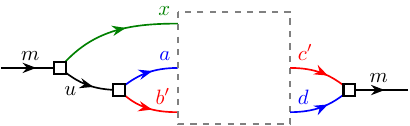}
\end{equation}
from the left and right, respectively. The resulting condition is that for all $a,d \in \mathcal{D}$, $b',c' \in \mathcal{D}'$, $u \in \mathrm{Irr}(\mathcal{F})$, $m \in \mathcal{D}\otimes \mathcal{D}'$, 
\be\begin{aligned}\label{eq:M-matrix-cond1}
&\sum_{e \in \mathcal{D}} \kappa_{ae} \, M_{eb}^{c'd}(m) \, \mathrm{G}^{(xab')m}_{eu}
 +
 \sum_{f' \in \mathcal{D}'} \kappa'_{b'f'} \, M_{af'}^{c'd}(m) \, \frac{\R{f'a}{m}}{\R{b'a}{u}} \, \mathrm{G}^{(xb'a)m}_{f'u} \\
 &= 
 \sum_{f' \in \mathcal{D}'} \kappa'_{f'c'} \, M_{ab'}^{f'd}(u) \, \mathrm{G}^{(xf'd)m}_{c'u}
 +
 \sum_{e \in \mathcal{D}} \kappa_{ed} \, M_{ab'}^{c'e}(u) \, \frac{\R{dc'}{m}}{\R{ec'}{u}} \, \mathrm{G}^{(xec')m}_{du} \ .
\end{aligned}\ee
An analogous computation for the second condition gives
\be\begin{aligned}\label{eq:M-matrix-cond2}
&\sum_{e \in \mathcal{D}} \wt\kappa_{ae} \, M_{eb'}^{c'd}(m) \, \frac{\R{b'e}{m}}{\R{b'a}{u}} \, \F{b'ax}{m}{eu}
+
\sum_{f' \in \mathcal{D}'} \wt\kappa'_{b'f'} \, M_{af'}^{c'd}(m) \, \F{ab'x}{m}{f'u}
\\
&= 
\sum_{f' \in \mathcal{D}'} \wt\kappa'_{f'c'} \, M_{ab'}^{f'd}(u) \, \frac{\R{dc'}{m}}{\R{df'}{u}} \, \F{df'x}{m}{c'u}
+
\sum_{e \in \mathcal{D}} \wt\kappa_{ed}  \, M_{ab'}^{c'e}(u) \, \F{c'ex}{m}{du}
\ .
\end{aligned}\ee
A special case of these conditions was already given in \cite{Runkel:2010ym}.

\subsubsection{Mutually commuting defects via a fusion graph}\label{sec:comm-via-fusion-graph}

For the second criterion we specialise to the case $\cD=\cD'=a$, i.e.\ both defects are equal and given by the elementary defect $a$. We can now decompose the identity on $\mathcal{E} := \cD \otimes \cD = a \otimes a$ as
\begin{equation}
    \includegraphics[valign = c]{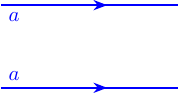}\quad = \sum_{m\in a \otimes a} \quad \includegraphics[valign = c]{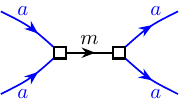}
\end{equation}
Write $\phi$ and $\ov\phi$ for the perturbing fields on $\mathcal{E}$.
Inserting the above decomposition of the identity between defect fields on $\cD \otimes \cD$, we obtain
\begin{equation}
    \phi = \sum_{m,n \in a \otimes a} \phi_{m \to n} ~~,\quad
    \ov\phi = \sum_{m,n \in a \otimes a} \ov\phi_{m \to n}  \ ,
\end{equation}
where, with $\wt\lambda  = \mu/\lambda$ and $\wt\lambda' = \mu/\lambda'$,
\begin{equation}\label{eq:fields-on-product}
\begin{split}
    \phi_{m\to n} &= \lambda \; \; \raisebox{0.21cm}{\includegraphics[valign = c]{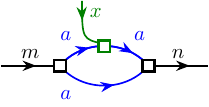}} + \lambda' \; \;  \raisebox{0.21cm}{\includegraphics[valign = c]{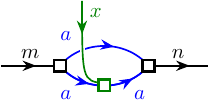}}\\
    &= \G{xaa}{n}{am} \left( \lambda + \lambda' \frac{\R{aa}{n}}{\R{aa}{m}} \right) \; \;  \raisebox{0.42 cm}{\includegraphics[valign = c]{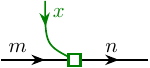}} \quad , \\
    \ov\phi_{m\to n} &= \frac{\mu}{\lambda} \;\; \raisebox{-0.24cm}{\includegraphics[valign = c]{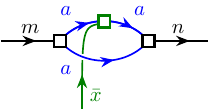}} + \frac{\mu}{\lambda'}  \;\; \raisebox{-0.24cm}{\includegraphics[valign = c]{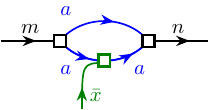}}\\
    &= \F{aax}{n}{am} \left( \frac{\mu}{\lambda'} + \frac{\mu}{\lambda}\frac{\R{aa}{n}}{\R{aa}{m}} \right) \;\; \raisebox{-0.36 cm}{\includegraphics[valign = c]{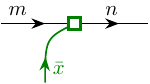}} \quad .
     \end{split}
\end{equation}
The prefactors of $\phi_{m\to n}$ and $\ov\phi_{m\to n}$ can be split into a part invariant under $\lambda \leftrightarrow \lambda'$ and a common factor that is not manifestly so,
\be\label{eq:phi-mn-prefactors}
\phi_{m\to n}~:~~  
\frac{\G{xaa}{n}{am}}{\R{aa}{m}} \Delta_{m\to n} 
\quad,\qquad
\ov\phi_{m\to n}~:~~ 
\frac{\mu}{\lambda\lambda'}\frac{\F{aax}{n}{am}}{\R{aa}{m}} \Delta_{m\to n}
\ ,
\ee
where
\be\label{eq:phi-mn-prefactors-nonsymmetric}
\Delta_{m\to n} = \lambda\R{aa}{m} + \lambda' \R{aa}{n} \ .
\ee

Now define a graph $\Gamma$ whose vertices are the labels $m \in a \otimes a$, and where there is an (undirected) 
edges between $m$ and $n$ if the prefactor of $\phi_{m \to n}$ is non-zero (which is equivalent to either of the prefactors of $\ov\phi_{m \to n}$, $\phi_{n \to m}$, or $\ov\phi_{n \to m}$ being non-zero by properties of the F-symbols). 

Consider a term in the perturbative expansion of $\mathcal{E}(\phi + \ov\phi)$ on a cylinder as in Figure~\ref{fig:perturbing term} (with $\mathcal{E}$ in place of $\cD$). Expand this expression further into a sum over elementary defects in $\mathcal{E}$, and expand the sum $\phi+\ov\phi$ at each insertion into $\phi_{m \to n}$ and $\ov\phi_{m \to n}$. Each non-zero term in this sum defines a closed path $\gamma$ on $\Gamma$ which is obtained by following the elementary defect labels $m \in \mathcal{E}$ as one passes along the defect from $s=0$ to $s=L$ on the cylinder. We also obtain a function $\delta$ from the edges of $\gamma$ to ${\pm 1}$ where $+1$ corresponds to an insertion of $\phi_{m \to n}$ and $-1$ to one of $\ov\phi_{m \to n}$.

Let $F_{\gamma,\delta}(\lambda,\lambda')$ be the product of the prefactors in \eqref{eq:phi-mn-prefactors}, where we take the prefactor of $\phi_{m\to n}$ if $\delta$ for that edge is $+1$ and of $\ov\phi_{m\to n}$ if it is $-1$. This now depends on the direction of $\gamma$ as e.g.\ the prefactors of $\phi_{m \to n}$ and $\phi_{n \to m}$ differ.
Then \eqref{eq:do-defects-commute} holds if for all closed paths $\gamma$ on $\Gamma$ and all maps $\delta$ from the edges of $\Gamma$ to $\{ \pm 1\}$ we have
$F_{\gamma,\delta}(\lambda,\lambda') = F_{\gamma,\delta}(\lambda',\lambda)$. But in order to verify this, we can drop the manifestly symmetric part in \eqref{eq:phi-mn-prefactors} and only consider the product of the common factor $\Delta_{m \to n}$ in \eqref{eq:phi-mn-prefactors-nonsymmetric}. This means we do not need the function $\delta$ and we define $F_\gamma(\lambda,\lambda')$ be the product of the factors $\Delta_{m \to n}$ as dictated by the path $\gamma$. 

Altogether, the second criterion is that \eqref{eq:do-defects-commute} holds if for all closed paths $\gamma$ on $\Gamma$ we have
\be \label{eq:F-inv-lamlam}
F_{\gamma}(\lambda,\lambda') = F_{\gamma}(\lambda',\lambda) \ .
\ee
We will see explicit examples of how to use this condition in Section~\ref{sec:charges-in-Vir}. Here we just make three general remarks which will be useful later:
\begin{itemize}
    \item For $m=n$ the factor $\Delta_{m \to m}$ is invariant under $\lambda \leftrightarrow \lambda'$. Hence in the graph $\Gamma$ and in the path $\gamma$ one may disregard edges from a vertex $m$ to itself.
    \item If $\gamma$ passes from $m$ to $n$ along an edge and later from $n$ to $m$, then $F_{\gamma}(\lambda,\lambda')$ will involve the product $\Delta_{m \to n}\Delta_{n \to m}$ which is invariant under $\lambda \leftrightarrow \lambda'$. Thus, if the are no non-trivial closed looks, i.e.\ if every path has to eventually retrace itself, then \eqref{eq:F-inv-lamlam} holds.
    \item If $\Gamma$ does contain loops, one needs to check for a choice of generating loops that the corresponding product of $\Delta_{m \to n}$'s is invariant under $\lambda \leftrightarrow \lambda'$. Combining this with the previous point, we again obtain that \eqref{eq:F-inv-lamlam} holds.
\end{itemize}

\subsection{The category of rigidly translation invariant defects}\label{sec:rigid-trans}

We briefly collect some properties of the category $\mathcal{T}(\mu)$ formed by rigidly translation invariant defects, and in particular of those obtained form joint perturbation of bulk and defect as above. Note that the CFT itself occurs as a special case for zero perturbation parameter.

\medskip

The objects in $\mathcal{T}(\mu)$ are rigidly translation invariant defects $\mathcal{X}$ in the CFT $\mathcal{C}(\mu \varphi)$ (or in any 2d\,QFT for that matter). 
Morphisms
$\mathcal{X} \to \mathcal{X}'$ in $\mathcal{T}(\mu)$ are defect changing fields which are topological in the sense that they can be moved freely on the defect line.
Note that the topological defects which still exist in $\mathcal{C}(\mu \varphi)$ form a full subcategory $\mathcal{T}_\mathrm{top}(\mu) \subset \mathcal{T}(\mu)$.

Most importantly, $\mathcal{T}(\mu)$ has a tensor product, given by fusion of defects in $\mathcal{T}(\mu)$. This is well-defined as by translation invariance no singularity arises in the fusion of parallel line defects.

The rigidly translation invariant defects $\cD(\lambda \psi + \mu/\lambda  \ov\psi)$ in $\mathcal{C}(\mu \varphi)$ obtained from solutions to the bulk commutation condition also form a full monoidal subcategory $\mathcal{T}_\mathrm{pert}(\mu) \subset \mathcal{T}(\mu)$. The category $\mathcal{T}_\mathrm{pert}(\mu)$ has a representation theoretic reformulation.  Namely, let $x \in \mathcal{F}$ label the perturbing field, and denote by $T(x) = \mathrm{1} \oplus x \oplus x^{\otimes 2} \oplus x^{\otimes 3} \oplus \cdots$ the tensor algebra of $x$ in (a completion of) $\mathcal{F}$.
Then $T(x)$ is a Hopf algebra in $\mathcal{F}$, and the bulk commutation condition for a line defect $\cD$ with defect fields $\psi,\ov\psi$ can be rephrased as the condition that $\cD$ carries the structure of a $T(x)$-$T(\overline x)$-Yetter-Drinfeld module (where the perturbing bulk field provides the Hopf-pairing), see \cite[Sec.\,4]{Buecher:2012ma}. The fusion of perturbed line defects amounts to the tensor product of Yetter-Drinfeld modules. Thus
\begin{quote}
    $\mathcal{T}_\mathrm{pert}(\mu)$ is monoidally equivalent to $T(x)$-$T(\overline x)$-Yetter-Drinfeld modules in $\mathcal{F}$.
\end{quote}
From this observation we see that $\mathcal{T}(\mu)_\mathrm{pert}$ is an abelian category (in particular it has kernels and quotients), and that it is rigid (has left and right duals).\footnote{
Note that while $T(x)$ only lives in a completion of $\mathcal{F}$ and does not have a dual, the object $\cD$ which carries the $T(x)$-$T(\overline x)$-Yetter-Drinfeld module structure is taken to be in $\mathcal{F}$, not in its completion. In particular, $\cD$ has a dual $\cD^*$ in $\mathcal{F}$, and indeed $\cD^*$ carries the structure of the dual Yetter-Drinfeld module. This situation is reminiscent of finite-dimensional representations of the universal enveloping algebra of a Lie algebra.
}
However, it is in general neither semisimple, nor spherical or pivotal (see~\cite[Sec.\,4.3]{Manolopoulos:2009np} for an example). It typically has continua of simple objects, and the fusion of two simple objects is generically again simple. In this sense, $\mathcal{T}_\mathrm{pert}(\mu)$ behaves a lot like evaluation representations of quantum affine algebras. This relation has been made precise in the perturbed free boson example treated in \cite{Buecher:2012ma}.

While the parameter $\mu \in \mathbb{C}$ enters the definition of $\mathcal{T}_\mathrm{pert}(\mu)$, it depends on $\mu$ only weakly in the sense that $\mathcal{T}_\mathrm{pert}(\mu)$ is monoidally equivalent (but not equal) to $\mathcal{T}_\mathrm{pert}(1)$ for all $\mu \in \mathbb{C}^\times$. However, in general $\mathcal{T}_\mathrm{pert}(0) \ncong \mathcal{T}_\mathrm{pert}(1)$.

\medskip

We propose that the monoidal categories
\begin{equation}
    \mathcal{T}(\mu)_\mathrm{top} ~\subset~
    \mathcal{T}(\mu)_\mathrm{pert} ~\subset~\mathcal{T}(\mu)
\end{equation}
contain important information about the family $\mathcal{C}(\mu \varphi)$ of deformed CFTs, and should all be considered together, instead of just $\mathcal{T}(\mu)_\mathrm{top}$.

\medskip

As a point in case, we expect that under the renormalisation group, defects in $\mathcal{T}(\mu)$ flow to topological defects in the infrared conformal field theory. Indeed, 
a defect $\mathcal{D}$ in a conformal field theory, which is both rigidly translation invariant and conformal, is necessarily 
topological. This is easy to see from the conditions on the defect operator $\widehat{\cD}$ on a circle. The conditions of conformality and translation invariance are, respectively,
\begin{align}
 [\, L_m - \ov L_{-m}, \widehat{\cD}\,] & = 0
 \comma \qquad \hbox{(all $m$)}\;,
 \label{eq:dhatconf}
 \\
 [\,L_0 + \ov L_0,\widehat{\cD}\,] & = 0
 \;.
 \label{eq:dhattrans}
 \end{align}
Taking \ref{eq:dhatconf} in the case $m=0$ and \eqref{eq:dhattrans}, we then get
\be
[\, L_0,\widehat{\cD}\,] = [\,\ov L_0,\widehat{\cD}\,] =0 
 \;.
 \label{eq:dhat0}
\ee
Now we consider the 
commutators
\begin{align}
 [\,L_0 , L_m - \ov L_{-m}\,] &=  -m \,L_m
 &&\Rightarrow\;
 [\,m\,L_m ,\widehat{\cD}\,] = 0
 &&\Rightarrow\;\;
 [\,L_m ,\widehat{\cD}\,]=0\comma &&\hbox{($m\neq  0$)}
 \label{eq:dhatlm}
 \\
 [\,\ov L_0 , L_m - \ov L_{-m}\,] &=  m \ov L_m
 &&\Rightarrow\;
 [\,m \,\ov L_{-m}, \widehat{\cD}\,] = 0
&&\Rightarrow\;\;
[\,\ov L_{-m}, \widehat{\cD}\,] = 0 \comma &&\hbox{($m\neq  0$)}
\label{eq:dhatlbarm}
\end{align}
and taking \eqref{eq:dhat0} with \eqref{eq:dhatlm} and \eqref{eq:dhatlbarm}, we see that $\widehat{\cD}$ satisfies the conditions of topological invariance:
\begin{align}
[\,L_m,\widehat{\cD}\,] = [\,\ov L_m,\widehat{\cD}\,] = 0
\comma \qquad \hbox{(all $m$)}
\;.
\label{eq:dhattop}
\end{align}

As a second application, consider two perturbed CFTs 
$\mathcal{C}(\mu \varphi)$ and 
$\mathcal{C}'(\mu' \varphi')$. Then the category of \textit{all} interfaces between $\mathcal{C}(\mu \varphi)$ and 
$\mathcal{C}'(\mu' \varphi')$ (not necessarily topological, or rigidly translation invariant, or conformal) is a bimodule category over $\mathcal{T}(\mu)$ and $\mathcal{T}'(\mu')$. For example, if $\mathcal{C}'$ is trivial, these interfaces are just boundaries for $\mathcal{C}(\mu \varphi)$, and fusing with defects in $\mathcal{T}(\mu)$ can be used to find boundary flows from defect flows and vice versa, see \cite{Constantin_Bachas_2004,Kormos:2009sk,Brunner:2009zt,Runkel:2010ym}.

\section{Non-local commuting charges in minimal models}\label{sec:charges-in-Vir}

Here we investigate the bulk commutation condition and the question of mutual commutativity of perturbed defects in the case of A-type minimal models $M(p,q)$. We first specialise the results in the previous section and review our conventions for minimal models. Then we treat the perturbations by $(1,2)$, $(1,3)$, $(1,5)$ and $(1,7)$ in $M(p,q)$ some detail, as well as all perturbations of $M(3,10)$. This section concludes with some further observations for which we do not provide a systematic treatment.

\subsection[ Properties of elementary defects \texorpdfstring{$\mathcal{D}$}{D} and of \texorpdfstring{$\mathcal{D}=\mathbf{1} \oplus x$}{D = 1 + x}]{\boldmath Properties of elementary defects \texorpdfstring{$\mathcal{D}$}{D} and of \texorpdfstring{$\mathcal{D}=\mathbf{1} \oplus x$}{D = 1 + x}}

\subsubsection{Elementary defect \boldmath $\mathcal{D}$}

We start by restating the bulk commutation condition \eqref{eq:commutation-condition-basis} in the case that $\mathcal{D}=a$ is an elementary defect: for all $b \in x \otimes a$ we must have
\begin{equation}\begin{split}\label{eq:one-defect-comm-cond}
b=a :&~~
	\left( \R{xa}{a} - \frac{1}{\R{ax}{a}}  \right) \F{ax\overline x}{a}{1a} =  \big( 1 - \F{xa \overline x}{a}{aa} \big)  \, \kappa \wt\kappa \ ,
\\
b \neq a :&~~
	\left( \R{xa}{b} - \frac{1}{\R{ax}{b}}  \right) \F{ax\overline x}{a}{1b} =  - \F{xa \overline x}{a}{ab} \, \kappa \wt\kappa  \ .
\end{split}\end{equation}
Note that by construction $a \in x \otimes a$, so that $b=a$ is indeed one of the values we need to consider.
If $\F{xa \overline x}{a}{aa} \neq 1$, if there is a solution at all, it must be given by
\begin{equation}\label{eq:one-defect-comm-cond-sol}
\kappa \wt\kappa =
	\left( \R{xa}{a} - \frac{1}{\R{ax}{a}}  \right) \frac{\F{ax\overline x}{a}{1a}}{1 - \F{xa \overline x}{a}{aa}}  \ .
\end{equation}
If $\F{xa \overline x}{a}{aa} = 1$ there can only be a solution if the left hand side of the $b=a$ condition is also zero.

\subsubsection{Elementary defect \boldmath $\mathcal{D}$ and $x \otimes x \cong \mathbf{1}$}\label{sec:x-invertible-with-fixedpoint}

To avoid having to treat the case $x \otimes x \cong \mathbf{1}$ separately in the examples below, we cover it here. Note that this condition implies that $x$ is invertible under fusion with the inverse being the dual, so that also $\overline x = x$. For $x$ to live on the defect $a$ we furthermore need $x \otimes a = a$. 

In \eqref{eq:one-defect-comm-cond} only the case $b=a$ remains. 
In Appendix~\ref{app:F-matrix-xx=1} we explain that
\begin{equation}
	\F{xax}{a}{aa} = \R{xx}{1} \in \{ \pm 1 \} \ .
\end{equation}
The other two possibilities $\R{xx}{1} \in \{ \pm i \}$ are excluded by the existence of a fixed point (see again the appendix).
Let us discuss the two cases in turn:
\begin{itemize}
    \item $\F{xax}{a}{aa} = -1$: In this case, \eqref{eq:one-defect-comm-cond} is uniquely solved by \eqref{eq:one-defect-comm-cond-sol}
    \item $\F{xax}{a}{aa} = +1$: Here \eqref{eq:one-defect-comm-cond} has a solution if and only if $\R{xa}a \, \R{ax}a = 1$. If this equality holds, \eqref{eq:one-defect-comm-cond} becomes $0=0$ and there are no constraints on $\kappa \wt\kappa$.
\end{itemize}

For minimal models we list R-matrices in the next section, and one finds that the condition $\F{xax}{a}{aa} = \R{xx}{1} =-1$ is equivalent to the conformal weight satisfying $h_x \in \mathbb{Z} + \frac12$, and $\F{xax}{a}{aa} = +1$ is equivalent to $h_x \in \mathbb{Z}$. In the latter case, $\R{xa}a \R{ax}a = 1$ holds automatically.

\subsubsection{Defect \boldmath $\mathcal{D}$ of the form $\mathbf{1} \oplus x$} 
\label{sec:xx=1+y-solution}

Next we turn to the case that $\mathcal{D} = \mathbf{1} \oplus x$, where $x$ is the label of the perturbing field. We assume that $x = \ov x$ and that $\F{xxx}{x}{11} \neq 1$.
For the application below we are interested in the case that $\psi$ and $\ov\psi$ do not have a component $x \to x$, that is, we make the additional assumption that
\begin{equation}\label{eq:D=1x-lamxx=0}
    \kappa_{xx} = 0 = \wt\kappa_{xx} \ .
\end{equation}
In this case one quickly checks that \eqref{eq:commutation-condition-basis} is equivalent to the equalities
\begin{equation}\label{eq:1+x-comm-cond}
    \kappa_{1x} \wt\kappa_{x1} = \kappa_{x1} \wt\kappa_{1x} =
    \frac{\R{xx}{1} - (\R{xx}{1})^{-1} }{(\F{xxx}{x}{11})^{-1}-1}
    = \frac{1}{\R{xx}{b}} - \R{xx}{b} \ , 
\end{equation}
where the last equality has to hold for all $b \in x \otimes x$, $b \neq \mathbf{1}$.
Here we used that necessarily\footnote{
\label{fn:F1x-nonzero}
This is a special case of the general observation that $\G{k\overline kj}{j}{1i} \neq 0$ if $j \in k \otimes i$, see e.g.\ \cite[Eqn.\,(2.60)]{Fuchs:2002cm} and observe that the left hand side there must be non-zero. By (2.61) in that paper, also $\F{j\overline kk}{j}{1i} \neq 0$.
}
$\F{xxx}{x}{1b} \neq 0$ for all $b \in x \otimes x$.
The F-matrix element in the above formula is given by
\begin{equation}\label{eq:Frob-Schur-dim}
    \F{xxx}{x}{11} = \frac{\nu_x}{\dim x} \ ,
\end{equation}
where $\nu_x \in \{ \pm 1 \}$ is the Frobenius-Schur indicator of $x = \ov x$, see \cite[Eqns.\,(2.20), (2.47)]{Fuchs:2002cm}.
Thus our assumption $\F{xxx}{x}{11} \neq 1$ is equivalent to $\dim x \neq \nu_x$.

\medskip

We show in Appendix~\ref{app:2-channel-fusion-proofs} that if $x \otimes x \cong 1 \oplus y$, i.e.\ if the decomposition of $x \otimes x$ contains precisely two simple objects, then \eqref{eq:1+x-comm-cond} automatically also holds for $b=y$. We also show there that \eqref{eq:M-matrix-cond1} and \eqref{eq:M-matrix-cond2} have a solution. Altogether:
\begin{quote}
Suppose that $x \in \mathrm{Irr}(\mathcal{F})$ satisfies $x = \ov x$ and $x \otimes x \cong 1 \oplus y$ for some simple object $y \in \mathrm{Irr}(\mathcal{F})$. Then the bulk commutation condition can be solved for $\mathcal{D} = 1 \oplus x$ with perturbation $x$, and the resulting perturbed defects mutually commute for different values of $\lambda$.
\end{quote}

\subsection{Chiral data for minimal models}

For $p,q \ge 2$ and coprime, we denote by $M(p,q)$ the A-type minimal model of central charge \be c=1-\frac{6(p-q)^2}{pq}\period\ee We write
\begin{equation}
    t = \frac{p}{q} \ .
\end{equation}
The irreducible representations of the Virasoro algebra that make up the model are indexed by Kac-labels
\begin{equation}
   \mathcal{I}_{p,q} := \big\{\, (r,s) \,\big| \, 1 \le r < p ~,~ 1\le s < q \, \big\} / \sim
   ~~ , \quad \text{where} ~~ (r,s) \sim (p-r,q-s) \ .
\end{equation}
The conformal weight of the primary state in the representation $(r,s)$ is
\begin{equation}
    h_{r,s} = \frac{(r-st)^2-(1-t)^2}{4t} \ .
\end{equation}
The vacuum representation is labelled by $\mathbf{1}=(1,1)=(p-1,q-1)$.
We will only need the fusion rules for representations of the form $(1,s)$, and these are
\begin{equation}\label{eq:1s-fusion-rules}
	(1,r) \otimes (1,s) ~\cong \hspace{-1em} \bigoplus_{\substack{u=|r-s|+1 \\\text{ step }2}}^{\min(r+s-1,2q-r-s-1)}
	\hspace{-1em} (1,u) \ .
\end{equation}
The R-matrix and twist are expressed in terms of the conformal weights as, for $a,b,c \in \mathcal{I}_{p,q}$,
\begin{equation}\label{eq:mm-theta-R}
    \theta_a = e^{2 \pi i h_a}
    ~~,\quad
    \R{ab}{c} = e^{\pi i (h_c-h_a-h_b)} \ .
\end{equation}
The quantum dimension of $a = (r,s) \in \mathcal{I}_{p,q}$ is given in terms of the modular S-matrix by $\dim(a) = S_{a1}/S_{11}$. Explicitly,
\be
  \dim(r,s) =  
  (-1)^{r-1} \, \frac{\sin( \pi r / t )}{\sin( \pi / t )} \cdot
  (-1)^{s-1} \, \frac{\sin( \pi s t)}{\sin( \pi t)} \ .
\label{eq:mm-dim}
\ee
The F-matrix entries can be determined via a closed form expression \cite{Dotsenko:1984ad,Furlan:1989ra} (summarised in \cite[App.\,A.1.1]{Graham:2001tg}), or recursively \cite{Runkel:1998he}. 
Here we will use the expressions from \cite{Runkel:1998he,Graham:2001tg}, the relation to the notation used there is
\begin{equation}
    \F{jkl}{i}{pq}
    ~=~
    \mathrm{F}_{pq}\big[ \begin{smallmatrix} j & k \\ i & l \end{smallmatrix} \big] \ .
\end{equation}
The inverse G of the F-matrix is given by
(combine \cite[Eqn.\,(2.61)]{Fuchs:2002cm} with \eqref{eq:mm-theta-R})
\be
  \mathrm{G}^{(abc)d}_{pq} = \F{cba}{d}{pq}  ~.
\label{eq:mm-GF}
\ee
The F-matrix has the symmetries
\be
  \F{abc}{d}{pq}
  = \F{bad}{c}{pq}
  = \F{cda}{b}{pq}
  = \F{dcb}{a}{pq} \ ,
\ee
and obeys (for $d \in b \otimes c$)
\be
  \F{1bc}{d}{db}
  = \F{b1d}{c}{db}
  = \F{cd1}{b}{db}
  = \F{dcb}{1}{db} = 1
\qquad \text{and} \qquad
  \F{aaa}{a}{11} = \frac{1}{\dim(a)}  \ .
\label{eq:F1p-vs-Fp1}
\ee
Comparing to \eqref{eq:Frob-Schur-dim} we see that the Frobenius-Schur indicators are $\nu_a = 1$ for all $a \in \mathcal{I}_{p,q}$.

\subsection[(1,2)-perturbation of the defect (1,1)\texorpdfstring{$\oplus$}{+}(1,2)]{\boldmath (1,2)-perturbation of the defect (1,1)\texorpdfstring{$\oplus$}{+}(1,2)}\label{sec:12-on-11+12}

That the bulk commutation condition can be solved for the $(1,2)$-perturbation and $\mathcal{D} = (1,1) \oplus (1,2)$ was already shown in Section~\ref{sec:xx=1+y-solution}, as was the fact that the resulting defect operators mutually commute. These solutions exist for all $M(p,q)$ with $q\ge 4$.

For concreteness, let us also solve \eqref{eq:1+x-comm-cond} explicitly. The conditions are
\be\begin{split}
	\kappa_{1x} \wt\kappa_{x1} = \kappa_{x1} \wt\kappa_{1x} 
	&=
	-2i \, \frac{\sin\!\big(\frac\pi2 t(x^2-1))}{\frac{\sin( \pi t x)}{\sin (\pi t)}+(-1)^x}\\
	&\overset{(*)}=
	-2i (-1)^x \sin\!\Big( \tfrac\pi4 \big(t(1+b^2-2x^2)-2(b+1)\big) \Big) \ ,
\end{split}
\ee
where $(*)$ is the consistency condition to check, and it has to hold for all $b \in x \otimes x$, $b \neq (1,1)$. For $x = (1,2)$ this is solved with
\begin{equation}
		\kappa_{1x} \wt\kappa_{x1} = \kappa_{x1} \wt\kappa_{1x} 	
		= -2 i \sin\!\big( \tfrac{\pi t}4 \big) \ .
\end{equation}

We note in passing that also $x=(1,3)$ solves the above condition, in that case one has to check $b=(1,3)$ and $b=(1,5)$. Thus the bulk commutation condition can be solved on $(1,1) \oplus (1,3)$ for the $(1,3)$-perturbation in such a way that $\psi$ and $\ov\psi$ do not involve a component $(1,3) \to (1,3)$.

\subsection{(1,3)-perturbation of the defect (1,2)}

Next we investigate the case where bulk and defect are perturbed by $(1,3)$ fields in $M(p,q)$ with $q \ge 4$. 
For the topological defect $\mathcal{D}$ we choose $\mathcal{D}=(1,2)$. 
This case was already treated in \cite{Runkel:2010ym}, and we review this result in the notation of the present paper.
We need to solve \eqref{eq:one-defect-comm-cond} for $x=(1,3)$, $a = (1,2)$, and $b\in (1,2) \otimes (1,3) \cong (1,2) \oplus (1,4)$. The case $b = (1,4)$ only occurs for $q \ge 5$. 

For $q=4$ we have $x \otimes x \cong \mathbf{1}$ which was treated in Section~\ref{sec:x-invertible-with-fixedpoint}. Since $h_{1,3} = 2t-1 = \frac12 p - 1$, and since $p$ has to be odd, we always have $h_{1,3} \in \frac12 + \mathbb{Z}$. Hence the bulk commutation condition can be solved for all $M(p,4)$.

Now suppose $q \ge 5$. The F-matrix entries occurring in \eqref{eq:one-defect-comm-cond} are
\be\begin{aligned}
    \F{233}{2}{12}&= 
    \frac{\Gamma(2 - 2 t) \Gamma(-1 + 3 t)}{\Gamma(1 - t) \Gamma(2 t)}   \ ,
&    
    \F{233}{2}{14}&= 
    -\frac{\sin(4 \pi t)}{\sin(3 \pi t)}
    ~
    \frac{\Gamma(2 - 2 t) \Gamma(-1 + 4 t)}{\Gamma(1 - t) \Gamma(3 t)}  \ ,
    \\
\F{323}{2}{22}&= -\frac{\sin(\pi t)}{\sin(3\pi t) }
 \ ,
&
        \F{323}{2}{24}&=     -\frac{\sin(4\pi t)}{\sin(3\pi t)}
    \,
    \frac{ \Gamma(2 - 3 t) \Gamma(-1 + 4 t) }{
\Gamma(1 - 2 t) \Gamma(3t)}
 \ .
\end{aligned}\ee
Using these, one can verify that \eqref{eq:one-defect-comm-cond} is uniquely solved by the value for $\kappa \wt\kappa$ given in \eqref{eq:one-defect-comm-cond-sol}:
\begin{equation}
    \kappa \wt\kappa = i \,
    \frac{\sin(3 \pi t)}{\cos(\pi t)}\,
    \frac{\Gamma(2 - 2 t) \Gamma(-1 + 3 t)}{\Gamma(1 - t) \Gamma(2 t)}
    \ .
\end{equation}
For the further examples discussed in Section~\ref{sec:charges-in-Vir} we will no longer list the explicit F-matrices or the explicit solution for $\kappa \wt\kappa$ -- one just obtains longer expressions of a similar form as above. Will will, however, discuss in detail when such solutions exist.

\medskip

That the defect operators obtained by perturbing the $(1,2)$ defect by $(1,3)$-fields mutually commute can be checked with the condition in Section~\ref{sec:comm-via-fusion-graph}.
Since $\mathcal{D} \otimes \mathcal{D} \cong (1,1) \oplus (1,3)$, the graph $\Gamma$ takes the simple form:
\begin{equation}
    \Gamma = \raisebox{-0.31cm}{\includegraphics[valign = b]{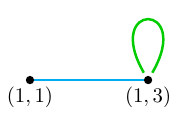}}
\end{equation}
This graph has no non-trivial closed loops, and so by the observations at the end of Section~\ref{sec:comm-via-fusion-graph}, the perturbed defect operators mutually commute for different values of $\lambda$.

\subsection{(1,5)-perturbation of the defect (1,3)}\label{sec:15-on-13}

Consider $M(p,q)$ for $q \ge 6$ perturbed by $x=(1,5)$, and the elementary topological defect $\mathcal{D}=a=(1,3)$. 

The case $q=6$ is the one treated in Section~\ref{sec:x-invertible-with-fixedpoint}. We have $h_{1,5} = 6t-2 = p-2$. Thus for the present $x$ and $a$, in $M(p,6)$ the commutation relation is satisfied for any choice of $\kappa \wt\kappa$.

Assume now that $q\ge 7$. We first check the value of $\F{xa \overline x}{a}{aa}$ in the case $b=a$ in \eqref{eq:one-defect-comm-cond} to see if it can be 1. Explicitly,
\begin{equation}
\F{535}{3}{33} = \frac{\sin(2 \pi t)}{4 \sin(5 \pi t) \cos(\pi t) \cos(2 \pi t)} =: f(t) \ .
\end{equation}
The only solutions to the trigonometric equation $f(t)=1$ are of the form $t \in \tfrac{s}{6} + 2 \mathbb{Z}$ for certain values of $s$. But for us $t = \frac{p}{q}$ with $p,q$ coprime and $q \ge 7$,
so that we can conclude 
\begin{equation}
\F{535}{3}{33} \neq 1 \quad \text{for $q \ge 7$ }.
\end{equation}
The case $b=a$ in \eqref{eq:one-defect-comm-cond} thus dictates the value for $\kappa \wt\kappa$ as given in $\eqref{eq:one-defect-comm-cond-sol}$. Since $x \otimes a = (1,5) \otimes (1,3) \cong (1,3) \oplus (1,5) \oplus (1,7)$ (where $(1,7)$ is only present for $q \ge 8$), the cases $b \neq a$ in \eqref{eq:one-defect-comm-cond} have to be checked for $b=(1,3)$ and $b=(1,7)$ (for $q \ge 8$), and we did verify that they indeed hold for the value of $\kappa \wt\kappa$ in $\eqref{eq:one-defect-comm-cond-sol}$

\medskip

The perturbed defect operators mutually commute by the criterion in Section~\ref{sec:comm-via-fusion-graph}.
Namely, the fusion of $\mathcal{D}$ with itself is $\mathcal{D} \otimes \mathcal{D} \cong (1,1) \oplus (1,3)\oplus (1,5)$. 
The graph $\Gamma$ depends on the value of $q$:
\be\begin{split}
q=6 ~:& \quad \Gamma =\raisebox{-0.53cm}{\includegraphics[valign = b]{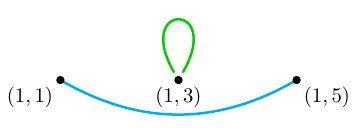}}
\\
q=7 ~:& \quad \Gamma = \raisebox{-0.53cm}{\includegraphics[valign = b]{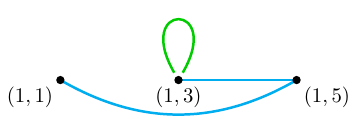}}
\\
q\ge 8 ~:& \quad	\Gamma = \raisebox{-0.53cm}{\includegraphics[valign = b]{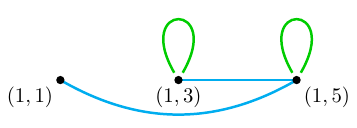}}
\end{split}\ee
In all these cases, the graph has no non-trivial closed loops, and so by Section~\ref{sec:comm-via-fusion-graph}, the perturbed defect operators mutually commute for different values of $\lambda$. 

\medskip

In summary, the bulk commutation condition for the $(1,5)$-perturbation on the $(1,3)$-defect can be solved for all $M(p,q)$ with $q \ge 6$, and in all cases the perturbed defect operators mutually commute for different values of $\lambda$.

\subsection{(1,7)-perturbation of the defect (1,5)}
\label{ssec:17on15}

Consider now $M(p,q)$ perturbed by $x = (1,7)$, and the elementary defect $\cD = a = (1,5)$. The perturbation $x$ is supported on the defect  $a$ only for $q\geq 9$, so that we restrict to those values in the following.  

\paragraph{Bulk commutation relation.} 
One may verify that 
\be
\F{757}{5}{55} -1 = -\frac{4 \cos (4 \pi  t)+4 \cos (8 \pi  t)+(\cos (4 \pi  t))^{-1}}{4 \cos (2 \pi  t)+4 \cos (4 \pi  t)+4 \cos (8 \pi  t)+2} \neq 0  \quad \text{for all} ~ t\in \mathbb{R}
\ ,
\ee
so that, if  a solution to the equation with $b=a$ in \eqref{eq:one-defect-comm-cond}  exist, then 
$\kappa\wt\kappa$ is given by  \eqref{eq:one-defect-comm-cond-sol}. To see if such a solution exist we must check that all the equations in \eqref{eq:one-defect-comm-cond} with $ a \otimes x \ni b \neq a$  are also satisfied. The content of $a \otimes x$ depends on  $q$ via \eqref{eq:1s-fusion-rules}. For $q \geq 9$ we have $(1,3), (1,5) \in a \otimes x$, and therefore \eqref{eq:one-defect-comm-cond} with $b = (1,3)$ must be satisfied for all choices of $q \ge 9$. 

We first check that the F-matrix element multiplying $\kappa\wt\kappa$ in the second line of \eqref{eq:one-defect-comm-cond} is not zero. Explicitly,
\be
  \F{757}{5}{53}  = \frac{4^{1-8 t} \sin (2 \pi  t) \tan (3 \pi  t) \, \Gamma\!\left(\frac{3}{2}-4 t\right) \Gamma (1-3 t) \Gamma (1-t) \Gamma (4 t)}{\sin (5 \pi  t) \sin (7 \pi  t)\Gamma (2-5 t) \Gamma (2-4 t) \Gamma (t) \Gamma\!\left(4 t-\frac{1}{2}\right)} \ .
\ee
For $t = \frac{p}{q}$ with $p,q$ coprime and $q \ge 9$, none of the $\Gamma$-function arguments is in $\mathbb{Z}_{\le 0}$, none of the $\sin$-arguments is in $\pi\mathbb{Z}$, and the $\tan$-argument is not in $\frac{\pi}2\mathbb{Z}$. So $\F{757}{5}{53}$ is well-defined and non-zero.

Next we substitute the value of $\kappa\wt\kappa$ in \eqref{eq:one-defect-comm-cond-sol} into the second line in \eqref{eq:one-defect-comm-cond}
with $b = (1,3)$. 
\be\label{eq:b13-case}
\left(\R{75}{3} - \frac{1}{\R{75}{3}}\right) \F{577}{5}{13}
~=~ - \F{757}{5}{53}
\left(\R{75}{5} - \frac{1}{\R{75}{5}}\right)\, \frac{\F{577}{5}{15}}{1-\F{757}{5}{55}}  \ .
\ee
The F-matrix elements are all non-zero (see also Footnote~\ref{fn:F1x-nonzero}), and 
$\R{75}{3} - 1/\R{75}{3} = -2 i \sin(16 \pi t)$. 

We first treat the case that $t \notin \frac{1}{16} \mathbb{Z}$. Then the left hand side is non-zero, and we can divide by it. Substituting the explicit values gives
\be \label{eq:cond73}
\frac{2+(\cos (8 \pi  t))^{-1}}{4 \cos (4 \pi  t)+4 \cos (8 \pi  t)+4 \cos (12 \pi  t)+6}  = 1 \ ,
\ee
whose solutions are $t \in f + \frac12\mathbb{Z}$, $f= \pm \frac15,\pm \frac19,\pm \frac29,\pm \frac1{10},\pm \frac1{18}$.
From this one can read off that the solutions with $q \ge 9$ are precisely 
\be 
t = \frac{p}{q}\comma\qquad  q = 9, 10, 18 \comma\qquad  \text{any $p$ coprime with $q$} \ .
\ee
The further equations depend on the values of $q$:
\be \begin{split}
q = 9&: \qquad a \ot x = (1,3) \oplus a \\
q = 10&:\qquad a \ot x = (1,3) \oplus a\oplus (1,7) \\
q = 18&: \qquad a \ot x = (1,3) \oplus a \oplus (1,7) \oplus (1,9)
\end{split}
\ee
We have verified that in each respective case the corresponding equations in \eqref{eq:one-defect-comm-cond} are satisfied by $\kappa\wt\kappa$ in \eqref{eq:one-defect-comm-cond-sol}.  

Next we turn to the case $t \in \frac{1}{16} \mathbb{Z}$. In order to solve \eqref{eq:b13-case}, also the right hand side must be zero, that is, we need $\R{75}{3} - 1/\R{75}{3} = 2 i \sin(12 \pi t) = 0$. This is the case for $t \in \frac{1}{48} \mathbb{Z}$. However, one can verify that then \eqref{eq:one-defect-comm-cond} does not hold for the choice $b=(1,7)$. 

Altogether, we find that the bulk commutation relation for the $(1,7)$-perturbation on the $(1,5)$-defect is satisfied only in the minimal models
\be 
M(p,9)\comma\qquad M(p,10)\comma \qquad M(p,18) \ .
\ee

\begin{figure}[t]
\begin{align*}
\allowdisplaybreaks
q = 9:& \qquad \Gamma = \raisebox{-.9cm}{\includegraphics[valign = b]{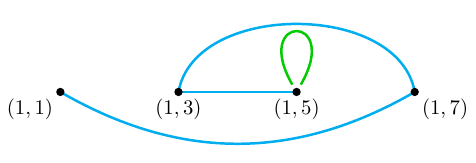} }\\
q = 10:&\qquad  \Gamma = \raisebox{-0.9cm}{\includegraphics[valign = b]{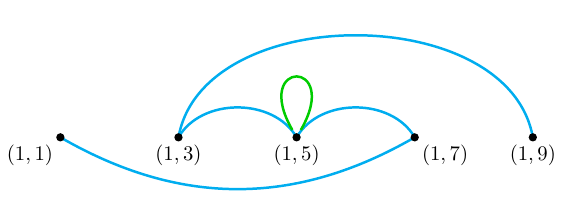}} \\
q = 18:& \qquad \Gamma = \raisebox{-2.05cm}{\includegraphics[valign = b]{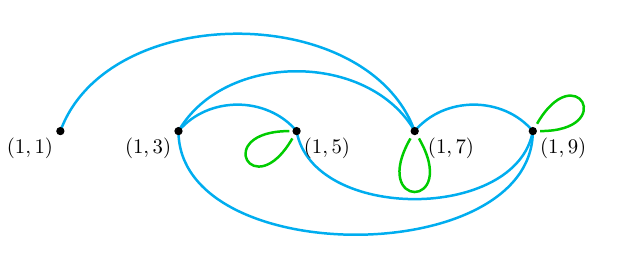}} 
\end{align*}
    \caption{The fusion graphs for the $(1,7)$-perturbation on of $(1,5)$-defect in $M(p,q)$.}
    \label{fig:17-fusion-graphs}
\end{figure}

\paragraph{Mutual commutation.}
Let us now discuss the mutual commutation. 
The fusion graphs $\Gamma$ for the three cases are listed in Figure~\ref{fig:17-fusion-graphs}.
Note that in writing the fusion graph for $q=10$ we omitted the edge $(1,3){-}(1,7)$. While this edge would be allowed by the fusion rules, the corresponding F-symbol
\be
\F{aax}{3}{a7}\eval_{q=10} \hspace{-0.5em}= 
\frac{\left(\sin\!\left(\frac{\pi  p}{5}\right)-
\sin\!\left(\frac{2 \pi  p}{5}\right)+\sin\!\left(\frac{3 \pi  p}{5}\right)-\sin\!\left(\frac{4 \pi  p}{5}\right)\right) 
\Gamma\!\left(2-\frac{3 p}{5}\right) \Gamma\!\left(\frac{4 p}{5}-1\right) \Gamma\!\left(\frac{2 p}{5}\right)}{\pi  \left(2 \cos\!\left(\frac{\pi  p}{5}\right)-1\right) \Gamma\!\left(\frac{3 p}{5}\right)} 
\ee
vanishes for any odd $p$ at $q=10$ (but not for $q=9,18$). 
Analogously, in the $q=18$ diagram, we did not draw the edge $(1,5){-}(1,7)$ as for $q=18$ (but not for $q =9,10$) 
\be
\F{aax}{5}{a7}\eval_{q=18} = 0
\ee
for any $p$ coprime to $q=18$.

\medskip

The argument in Section~\ref{sec:comm-via-fusion-graph} implies that the perturbed defects mutually commute for $q=9$ and $q=10$ since there are no non-trivial loops. 

For $q=18$ one needs to check that \eqref{eq:F-inv-lamlam} holds for a selection of generating loops $\gamma$. We choose
\be 
\gamma_{359}= 3\to5\to9\to3\comma \qquad \gamma_{379}= 3\to7\to9\to3 \ .
\ee 
Since the two differ only by $5\leftrightarrow 7$, we write generically $\gamma_{3k9}$, with $k = 5,7$. 
We need to check \eqref{eq:F-inv-lamlam} for these two path:
\be
F_{\gamma_{3k9}}(\lambda,\lambda') = F_{\gamma_{3k9}}(\lambda',\lambda) 
\quad \text{for} \quad k = 5,7 \ .
\ee
One readily computes that
\be
\begin{split}
&F_{\gamma_{3k9}}(\lambda,\lambda') - F_{\gamma_{3k9}}(\lambda',\lambda) 
\\
&\quad = \lambda\, \lambda' (\lambda-\lambda') (\R{55}{3} - \R{55}{k})(\R{55}{3} - \R{55}{9})(\R{55}{k} - \R{55}{9}) \ .    
\end{split}
\ee
For $q = 18$ one has
\be 
\R{55}{3}- \R{55}{9} = -e^{- i 5\pi  p/9} \left(1+e^{i \pi  p}\right) = 0\period
\ee
Altogether we see that \eqref{eq:F-inv-lamlam} holds for the two generating loops, and hence for all loops. By the criterion in Section~\ref{sec:comm-via-fusion-graph} we can conclude that also for $q=18$ the perturbed defect operators mutually commute.

\subsection[The minimal model \texorpdfstring{$M(3,10)$}{M(3,10)}  in detail]{\boldmath The minimal model \texorpdfstring{$M(3,10)$}{M(3,10)}  in detail}\label{sec:M310}

Up to now we have looked at a fixed combination of perturbation and defect and asked in which models the bulk commutation condition is satisfied, and whether the perturbed defects mutually commute. Here we want to instead consider one example, namely the minimal model $M(3,10)$, and look at two situations: perturbing an elementary defect $\mathcal{D}=a$ by some $x$, and perturbing $\mathcal{D} = \mathbf{1} \oplus x$ by $x$. In both cases it is understood that the bulk CFT is perturbed by the bulk field labelled $x$.

We note that the $M(3,10)$ minimal model is one of the three cases where a product of two minimal models is again a minimal model. Here, the D-invariant of $M(3,10)$ is the product of two copies of the Lee-Yang model $M(2,5)$. This and the other two cases are treated in more detail e.g.\ in \cite[App.\,C]{Quella:2006de}.

\subsubsection{\boldmath Non-local charges in $M(3,10)$ from $\mathcal{D}=a$}

\begin{table}
    \centering
\[
\begin{array}{c|l}
a & x \in a\otimes a \\ \hline
(1,1) & (1,1) 
\\
(1,2) & (1,1) \, , \, (1,3)
\\
(1,3) & (1,1) \, , \, (1,3) \, , \, (1,5) 
\\
(1,4) & (1,1) \, , \, (1,3) \, , \, (1,5)  \, , \,  (1,7)
\\
(1,5) & (1,1) \, , \, (1,3) \, , \, (1,5)  \, , \,  (1,7)  \, , \,  (1,9)
\\
(1,6) & (1,1) \, , \, (1,3) \, , \, (1,5)  \, , \,  (1,7)
\\
(1,7) & (1,1) \, , \, (1,3) \, , \, (1,5)
\\
(1,8) & (1,1) \, , \, (1,3)
\\
(1,9) &  (1,1)
\end{array}
\]
    \caption{The (anti)holomorphic defect fields on the elementary defect $a$ are indexed by those $x$ that appear in the fusion $a \otimes a$.}
    \label{tab:fusin-in-M310}
\end{table}

As unique representatives of the Kac-labels in $M(3,10)$ we can take $(1,s)$ with $1 \leq s \leq 9$.
The primary holomorphic and antiholomorphic defect fields on the elementary topological defect $\mathcal{D}=a$ are indexed by those $x$ that are contained in the fusion $a \otimes a$, see Table~\ref{tab:fusin-in-M310}.

For each $a$, and for each $x \in a \otimes a$ we ask if a solution to the bulk commutation condition exists, and if the criteria in Section~\ref{sec:def-op-commute} can be used to show that the perturbed defects mutually commute. The results are collected in Table~\ref{tab:M310-single-defect}.

Checking whether the bulk commutation condition holds for a given $a$ and $x$ in the table simply amounts to verifying whether \eqref{eq:commutation-condition-basis} has a solution in that case or not. We have checked this condition, with the results in the table.

Mutual commutativity of the perturbed defects can be seen using the criterion in Section~\ref{sec:comm-via-fusion-graph}. Let us go through this criterion for the columns one by one:
\begin{itemize}
    \item $x=(1,3)$: In this case the fusion graph $\Gamma$ cannot have non-trivial closed loops.
    
    \item $x=(1,5)$: The case $a=(1,3)$ was treated in Section~\ref{sec:15-on-13}, and the case $a=(1,7)$ works analogously. For $a=(1,5)$ the fusion graph $\Gamma$ turns out to have no loops due to the vanishing of F-matrix elements:
    \be
\F{555}{5}{53} = 0 \quad \text{and} \quad \F{555}{5}{57} = 0 \ .
    \ee
    Thus the edges $(1,3){-}(1,5)$ and $(1,5){-}(1,7)$ which would by allowed by the fusion rules are not present (see \eqref{eq:fields-on-product}) and there are indeed no loops:
    \be
    \Gamma = \raisebox{-0.55cm}{\includegraphics[valign = b]{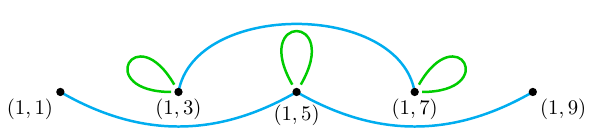}}
    \ee
Note that in this example, the composition of perturbed defect operators decomposes into a sum of a $(1,5)$-perturbation on $(1,1)\oplus(1,5)\oplus(1,9)$ and on $(1,3)\oplus(1,7)$, both of which therefore satisfy the bulk commutation condition.

\item $x=(1,7)$: This case was treated in Section~\ref{ssec:17on15}.

\item $x=(1,9)$: Note that $x \otimes x \cong \mathbf{1}$  and we are in the situation discussed in Section~\ref{sec:x-invertible-with-fixedpoint}. The conformal weight of the primary state in the representation $x$ is $h_x = 2$, and so \eqref{eq:one-defect-comm-cond} just becomes $0=0$, i.e.\ the bulk commutation condition is solved for any choice of $\kappa$ and $\wt\kappa$. The fusion graph $\Gamma$ again has no loops:
    \be
    \Gamma = \raisebox{-0.5cm}{\includegraphics[valign = b]{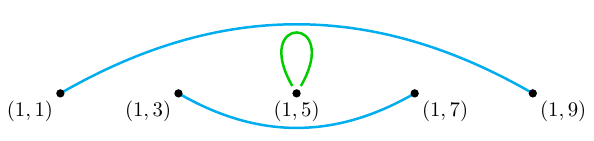}}
    \ee
\end{itemize}

\begin{table}
    \centering
\[
\begin{array}{c|cccc}
 \text{Defect, $a$} & \multicolumn{4}{c}{\text{Perturbation, $x$}}\\
    & (1,3) &  (1,5) & (1,7) & (1,9) \\
    \hline
(1,2) & \ckb & -& -& - \\   
(1,3) & \ckb & \ckb & -& - \\   
(1,4) & \ckb & \times & \times^* & - \\   
(1,5) & \ckb & \ckb 
& \ckb^* & 
\ckb^* 
\\   
(1,6) & \ckb & \times & \times^* & - \\   
(1,7) & \ckb & \ckb &- & - \\   
(1,8) & \ckb &- &- &-  \\   
\end{array}
\]
    \caption{Perturbation by $x$ on the topological defect $a$. The entries are as follows: $-$, the perturbation does not exist; $\times$, the perturbed defect does not commute with the Hamiltonian; $\ckb$, the non-local charges commute with the Hamiltonian and we can also prove that they mutually commute; ${}^*$ indicates that the non-local charges require regularisation and so the arguments are only formal. 
    }
    \label{tab:M310-single-defect}
\end{table}

Recall that the D-invariant of $M(3,10)$ is equivalent to two copies of $M(2,5)$. Denote the representations of $M(2,5)$ by $\mathbf{1}$ and $\phi$. Then in terms of $M(3,10)$-representations, the representations in the product decompose as
\be
\mathbf{1} \times \mathbf{1} = (1,1) \oplus (1,9)
~~,\quad
\mathbf{1} \times \phi = \phi \times \mathbf{1} = (1,5)
~~,\quad
\phi \times \phi = (1,3)\oplus(1,7) \ .
\ee
In $M(2,5)$, the $\phi$-perturbation on the $\phi$-defect satisfies the bulk commutation condition and mutual commutativity, and so it is maybe not too surprising to find the same in $M(3,10)$. But it would still be interesting to understand in more detail how these conditions behave under gauging of topological symmetries (a $\mathbb{Z}_2$ in this case).

\subsubsection{\boldmath Non-local charges in $M(3,10)$ based on two defects \texorpdfstring{$\mathbf{1} \oplus x$ }{1 + x}}

The perturbations that are supported on a superposition of defects $\mathbf 1\oplus a$ are those that are supported by the defect $a$ alone, and those which interpolate the defects $\mathbf 1$ and $a$, which is forced to be of type $a$. In Table~\ref{tab:M310-defect-1+x} we give the results for the defects of type $\mathbf 1\oplus x$ perturbed by $x$. This includes cases in which the defect $x$ supports the field $x$ and cases in which it does not.

\begin{table}
    \centering
\[
\begin{array}{c|cccc}
 \text{Defect, $1 \oplus x$} & \multicolumn{4}{c}{\text{Perturbation, $x$}}\\
    \hline
(1,1)\oplus (1,2) & \ckb\\   
(1,1)\oplus (1,3) & \ckb\\ 
(1,1)\oplus (1,4) & \times  \\   
(1,1)\oplus (1,5) & \cka \\   
(1,1)\oplus (1,6) & \times \\   
(1,1)\oplus (1,7) & \cka^*  \\   
(1,1)\oplus (1,8) & 
\ckb^*    \\ 
(1,1)\oplus (1,9) & \cka^*    \\ 
\end{array}
\]
    \caption{The $x$-perturbation on the defect $\mathcal{D} = \mathrm{1} \oplus x$. The entries in the right column are as in Table~\ref{tab:M310-single-defect}, and in addition
    a single tick $\cka$ means that the perturbed defect commutes with the Hamiltonian, but we are unsure if the charges mutually commute.
    }
    \label{tab:M310-defect-1+x}
\end{table}

Let us explain the entries for mutual commutativity in more detail. The case $x=(1,2)$ is treated in Section~\ref{sec:12-on-11+12}, relying on the criterion in Section~\ref{sec:xx=1+y-solution}. The $(1,3)$-perturbation of $(1,1)\oplus(1,3)$ can be obtained by taking the product of two $(1,2)$-defects perturbed by $(1,3)$. Since by Table~\ref{tab:M310-single-defect}, these do mutually commute, the same holds for their products. For $x=(1,8)$ we have $x \otimes x = (1,1) \oplus (1,3)$ and so the criterion in Section~\ref{sec:xx=1+y-solution} applies: both, the bulk commutation condition and mutual commutativity are satisfied.

\medskip

The representations $(1,s)$ of $M(3,10)$ with $s$ even do not occur in the product of two $M(2,5)$-models and whether or not they solve the bulk commutation condition can not be predicted from there. For these, we only find solutions on $(1,1) \oplus (1,s)$ for $s=2,8$. The $(1,2)$-perturbation is generically integrable and the $(1,8)$-perturbation is only formal as it would need regularisation.

\subsection{Further observations}
\label{ssec:furtherobs}

After the systematic treatment in the previous sections, here we collect some additional observations which indicate that we are still quite far from a complete picture of solutions to the commutation condition in minimal models:
\begin{itemize}

\item In Section~\ref{ssec:17on15} we saw that for the $(1,7)$-perturbation on the $(1,5)$-defect allows for a solution of the bulk commutation condition iff $q=9,10,18$. 
If we instead take the defect $(1,1) \oplus (1,7)$, there are solutions for $M(p,11)$ and $M(p,18)$.

\item We found solutions to the $(1,4)$-perturbation on $(1,1) \oplus (1,4)$ in $M(p,5)$,  $M(p,6)$ for all $p$, 
in $M(p,7)$ for all even $p$, 
and exceptionally in $M(2,9)$, $M(2,11)$.
For the $(1,6)$-perturbation on $(1,1) \oplus (1,6)$ we found solutions for $q=7,8,9$ (and exceptionally in $M(2,11)$). 

\item In addition to the perturbations already mentioned, we looked at the bulk commutation relation for $(1,2k+1)$-perturbations of elementary defects for $k=4,5,6$. 
Again, we do not have a complete picture for these perturbations, but have found the following.
\begin{itemize}

    \item[i)] The $(1,9)$-deformation of $(1,5)$ defect satisfies bulk commutation in the minimal models $M(p,10)$, $M(p,11)$, $M(p,12)$. 
    The same deformation on the $(1,6)$ defect is supported on  $M(p,11)$, $M(p,12)$, on the $(1,7)$ defects commutes in $M(p,12)$, $M(p,14)$.    
    We have not found any solutions for the $(1,9)$ deformation of the elementary defects $(1,n)$, for $8 \leq n \leq 19$.
    
    \item[ii)] The $(1,11)$-deformation of the $(1,6)$-defect  satisfies the bulk commutation in $M(p,12)$ and $M(p,13)$, 
and of the $(1,7)$-defect in $M(p,13)$ and $M(p,14)$. We did not find any other solution to the bulk commutation relation for the $(1,11)$-deformation of the $(1,n)$-defect for $8 \leq n \leq 19$.

    \item[iii)] The $(1,13)$-deformation of the $(1,7)$-defect satisfies the bulk commutation in $M(p,14)$, $M(p,15)$ and $M(p,16)$, of the $(1,8)$-defect in $M(p,15)$ and $M(p,16)$,
    and of the $(1,9)$-defect in
    $M(p,16)$, $M(p,18)$.  We have not found any solutions for the $(1,13)$ deformation of the elementary defects $(1,n)$, for $10 \leq n \leq 19$.
\end{itemize}

\item In all solutions to the bulk commutation condition that we found, the perturbing field is of the form $(r,s)$ with either $r=1$ or $s=1$.

\end{itemize}
We did not try to check mutual commutativity in these cases.

\section{Local conserved charges}\label{sec:local-charges}

It has long been recognised that the presence of non-trivial local conserved quantities can lead to the integrability of quantum field theories -- 
see e.g.\ \cite{Dorey:1996gd} for a discussion. 
Up to now, these have only been found in cases related to some classically integrable system (e.g.\ generalised affine Toda theories/Drinfeld-Sokolov hierarchies) or where there is a relation to  extended chiral symmetry.

For the Virasoro minimal models, the three standard integrable perturbations,
$(1,2)/(5,1)$, $(1,3)/(3,1)$ and $(1,5)/(2,1)$,
can be identified with quantisations of the affine Toda theories related to  $a_2^{(2)}$, $a_1^{(1)}$ and $a_2^{(2)}$ respectively.
It is known \cite{Feigin:1991qm} that there are generically (i.e.\ for irrational central charge $c$) local conserved quantities  with spins equal to the exponents of the corresponding affine algebra, that is all odd spins for $(1,3)/(1,3)$ perturbations, and all spins congruent to 1 or 5 mod 6 for the other perturbations. However, we do not know of a statement or proof of the general number of conserved quantities of each spin for each perturbation in the case of minimal models.
A conserved charge predicted in the classical theory may vanish in the quantum model (as happens for the spin $5$ charge for the $(1,2)$ perturbation of the Ising model), and the number of charges may also be increased if there is a relation to an extended symmetry (as happens for the $(1,3)$ perturbation of the Ising model where the perturbing field has weight $h=\frac12$ and there are extra conserved charges due to its relation to a free fermion).

It is also possible that a perturbation can be identified with a different integrable system, which can lead to further particular perturbations being integrable in specific models, or a change in the number of conserved quantities. As an example of the last case, the $(1,2)$ perturbation of the Ising model can also be identified with a quantised $e_8^{(1)}$ affine Toda theory, which ``explains'' why the spin $5$ charge vanishes as 5 is not an exponent of $e_8^{(1)}$.

In the series of papers 
\cite{Bazhanov:1994ft,Bazhanov:1996dr,Bazhanov:1996aq}, it was shown that there is a quantum version of the classical result that  the monodromy matrices have an asymptotic expansions in the local conserved charges. 
These papers defined ``quantum transfer matrices'' (which can be identified as perturbed defects) 
which have an asymptotic expansion over the set of local conserved charges -- i.e.\ there is an intrinsic relation between the non-local perturbed defects and the local charges. 

We have consequently two questions we can try to answer -- are there local conserved quantities for the new perturbations that we have found that are not related to known classical integrable hierarchies? -- if not, is there a reason why the perturbed defects we have defined do not have an asymptotic expansions over local charges?
We consider these two questions in Sections~\ref{ssec:localsearch}
and \ref{ssec:asymptotic} respectively.

\subsection{Search for local conserved quantities}
\label{ssec:localsearch}

The consideration of quantum local conserved charges in perturbed conformal field theories defined entirely through the properties of the quantum field theory (rather than derived from some underlying construction, such as the non-local charges that can be shown to exist through a free-field construction related to Toda theory) dates back to \cite{Zamolodchikov:1989hfa}. 
This states that the expression for the antiholomorphic derivative in the perturbed theory, $\overline\partial W$ of a field $W$ which is holomorphic in the unperturbed theory, can be expanded out order by order in $\mu$,
\begin{align}
    \overline\partial W = \sum_{n=1}
    (2i\mu)^n V_n= (2i\mu) V_1 + \sum_{n=2}
    (2i\mu)^n V_n\;.
    \label{eq:zampert}
\end{align}
Using $[\cdot]$ to denote (mass) scaling dimension as usual, 
$[\overline\partial W] = h_W + 1$ and
$[\mu] = 2 - 2 h_\varphi$ so 
$[V_n] = 2 n h_\varphi - 2n + h_W + 1$.
Since $[\varphi] = 2 h_\varphi$, $V_1$ is always an allowed term, as a level $h_W-1$ descendant of $\varphi$. Whether other $V_n$ are allowed or not depends in detail on the model and perturbation under consideration.

For a rational model, 
since the dimension of $\mu$ is rational, there will always be some power $N$ for which $\mu^N$ has integer dimensions, and so higher contributions $V_{1+rN}$ will always be possible for $h_W$ large enough). 

It is a necessary condition for the conservation of $W$ that $V_1$ is a total $\partial$-derivative.
In the lucky circumstance that the expansion \eqref{eq:zampert} terminates at the first term, this constitutes a proof of the existence of a conserved charge in the perturbed model. In general, if $V_1$ is a total derivative, the conservation condition may be obstructed by higher terms $V_n$, but if $V_1$ is {\em not} a total derivative, then this cannot be improved by higher order terms and this proves that there is no associated conserved current.
So, from now, on, we focus entirely on the condition on $V_1$. 

The condition on $V_1$ is given in Equation~\eqref{eq:pcond} in Appendix~\ref{app:localcharge} and  can be checked explicitly (using Mathematica) in any particular cases. Note that the existence of solutions to \eqref{eq:pcond} has no bearing on the mutual commutativity, or otherwise, of the corresponding charges.

The general expectation is that we will find solutions to \eqref{eq:pcond} with spins equal to those in the corresponding classical system whenever the perturbation can be identified as the quantisation of classical integrable system, although some of the charges may vanish through singular vector conditions in particular models, and extra solutions can also arise -- but  up to now nobody has identified any perturbed models outside the classically integrable class which have local conserved charges.

This is what we also find (in all but one case in which the classical system is so far unidentified). We have performed checks in a large number of models for a large number of perturbations and we have found solutions to \eqref{eq:pcond} in the following cases.

\subsubsection{\boldmath (1,2), (1,3) and (1,5)  perturbations}

These perturbations are are related to the $a_1^{(1)}$ (for (1,3)) and $a_2^{(2)}$ (for (1,2) and (1,5)) algebras and classically one has conserved charges of all odd spins in the first case and all spins congruent to $1$ or $5$ mod 6 in the second case. This is not exactly the case for the quantum systems since, as mentioned already, some charges may vanish unexpectedly and extra charges can occur, as with the examples in Table~\ref{tab:121315}.

{
\begin{table}[t]
\resizebox{\textwidth}{!}{  \renewcommand{\arraystretch}{1.2}
$
    \begin{array}{c|l|c|ccccccccccccccccc}
        M(p,q) & \text{pert}^{n.} & h & 4 & 5 & 6 & 7 & 8 & 9 & 10 & 11 & 12 & 13 & 14 & 15 & 16 & 17 & 18 & 19 & 20\\
        \hline
  (2,5)& (1,2) & -\frac{1}{5} & 
    - & - & 1  & - & 1 & - & - & - & 1 & - & 1 & - & - & - & 1 & - & 1
 \\
 & (1,3) & -\frac{1}{5} & 
    -^1 & - & 1  & - & 1 & - & -^1 & - & 1 & - & 1 & - & -^1 & - & 1 & - & 1
  \\ \hline
  (2,11)& (1,2) & -\frac{4}{11} & 
    - & - & 1  & - & 1 & - & - & - & - & - & 1 & - & - & - & 1 & - & 1 \\
 & (1,3) & -\frac{7}{11} & 
    1 & - & 1  & - & 1 & - & 1 & - & 1 & - & 1 & - & 1 & - & 1 & - &1 
 \\
 & (1,4) & -\frac{9}{11} & 
    - & - & -  & - & 1 & - & - & - & 1 & - & 1 & - & - & - & 1 & - & 1
 \\
 & (1,5) & -\frac{10}{11} & 
    - & - & 1  & - & 1 & - & - & - & 1 & - & 1 & - & - & - & 1 & - & 1
  \\ \hline
  (3,4)& (1,2) & \frac{1}{16} & 
    - & - & -^1 & - & 1 & - & - & - & 1 & - & 1 & - & - & - & 1 & - & 1
  \\
       & (1,3)^2 & \frac{1}{2} &
    1 & - & 1 & - & 1 & - & 1 & - & 2 & - & 2 & 1 & 3 & 1 & 4 & 2 & 5
       
       \\ \hline
(4,5) & (1,2) & \frac{1}{10} &
    - & - & 1 & - & 1 & - & - & - & 1 & - & 1 & - & - & - & 1 & - & 1 
       \\
       & (1,3) & \frac{3}{5} &
    1 & - & 1 & - & 1 & - & 1 & - & 1 & - & 1 & - & 1 & - & 1 & - & 1
       \\
       & (3,1)^2 & \frac{3}2 &
    1 & - & 1 & - & 2 & - & 2 & 1 & 3 & 1 & 5 & 2 & 6 & 4 & 9 & 6 & 13 
       \\
       &(2,1) & \frac{7}{16} &
  -  & - & 1 & - & 1 & - & - & - & 1 & - & 1 & - & - & - & 1 & - & 1 
         \\
       &(2,2) & \frac{3}{80} &
  -  & - & - & - & - & - & - & - & - & - & - & - & - & - & - & - & - 
       \\ 
\end{array}
$   }
\caption{Spins of conserved currents in selected models. In some cases, currents which are expected by relation to classical integrable systems vanish (marked ${}^1$); in some cases there are extra currents (marked ${}^2$). See the text for details.}
    \label{tab:121315}
\end{table}
}

\subsubsection{\boldmath (1,4) perturbations}

These are not typically identifiable with a classical integrable model, and in all the cases we have checked there are no local conserved charges, apart from one exception, the $(1,4)$ perturbation in $M(2,11)$. In this particular case the chiral algebra can be identified with the degeneration of the $WG_2$ chiral algebra to at $c=-232/11$ to the Virasoro algebra \cite{Milas:2023cwx}, and the $(1,4)$ perturbation can be identified as related to $d_4^{(3)}$ affine Toda theory. Again, see Table~\ref{tab:121315}.

\subsubsection{\boldmath (1,7) perturbations}
\label{sssec:17perts}

We have searched for solutions to the first-order conservation condition \eqref{eq:pcond} for $(1,7)$ perturbations of the models $M(p,q)$ with $p,q \leq 20$, which includes the specific cases $q\in\{9,10,18\}$ discussed in Section~\ref{ssec:17on15}. The results are in Table~\ref{tab:spins17}.
In almost all cases we found no conserved charges with spins $\leq 16$, with only the following exceptions:

\begin{itemize}
    \item $M(2,9)$ - the perturbation can be identified with $(1,2)$
    \item $M(2,11)$ - the perturbation can be identified with $(1,4)$ and so, as above, the  chiral algebra can be identified with $WG_2$ and the perturbation can be identified with $d_4^{(3)}$ affine Toda theory.
 
    \item $M(3,8)$ - the perturbation can be identified with $(2,1)$, and has extra conserved charges as $h=3/2$.
    \item $M(p,8)$ - the perturbation is a simple current of half-integer weight and is related to an extended chiral algebra. We conjecture that in each case there is a solution of \eqref{eq:pcond} of spin $(3p-7)$ and have checked that there is a conserved current of spin $20$ for $p=9$. The cases with $p>7$ do not appear in Table~\ref{tab:spins17} because this is greater than 16.
    These are all irrelevant perturbation and so there are an infinite number of extra terms which can arise in the 
    the conservation condition \eqref{eq:zampert}.
      
    \item $M(13,18)$ at $c=14/39$ - we do not have an identification, as yet, of this with the quantisation of a classical integrable system. This is, again, an irrelevant perturbation and so there are an infinite number of extra terms which can arise in the conservation condition \eqref{eq:zampert}.
   
\end{itemize}

This is not a proof of the existence of these conserved quantities since in all cases there are potential extra terms coming from $\lambda^3 V_3$ with $V_3$ in the $(1,5)$ representation, as well as possibly higher terms, and we do not know a way to calculate these potential corrections.

\begin{table}[t]
\renewcommand{\arraystretch}{1.2}
\[
    \begin{array}{l|c|c|ccccccccccccccccc}
        M(p,q) & c & h & 4 & 5 & 6 & 7 & 8 & 9 & 10 & 11 & 12 & 13 & 14 & 15 & 16 
        \\
        \hline
  (2,9)&-\frac{46}3& -\frac{1}{5} & 
    - & - & 1  & - & 1 & - & - & - & 1 & - & 1 & - & - 
 \\
  (2,11)&-\frac{232}{11}& -\frac{9}{11} & 
    - & - & -  & - & 1 & - & - & - & 1 & - & 1 & - & -
 \\
  (3,8)&-\frac{21}{4}& \frac{3}{2} & 
    - & - & 1  & - & 2 & - & 2 & 1 & 3 & 1 & 5 & 3 & 6 
 \\
  (5,8)&-\frac{7}{20}& \frac{9}{2} & 
    - & - & - & - & 1 & - & - & - & - & - & - & - & - 
 \\
  (7,8)&\frac{25}{28}& \frac{15}{2} & 
    - & - & -  & - & - & - & - & - & - & - & 1 & - & - 
 \\
   (13,18)&\frac{14}{39}& \frac{17}{3} & 
    - & - & -  & - & - & - & - & - & - & - & 1 & - & 1 &  
 \\
        \hline
    \end{array}
\]    \caption{Models $M(p,q)$ with $p,q \leq 20$ for which there are currents with spins $\leq 16$ for $\psi_{1,7}$ perturbations which solve the first-order conservation condition \eqref{eq:pcond}, together with the number of independent currents for each spin.}
    \label{tab:spins17}
\end{table}

\subsection{Local charges from perturbed defects}
\label{ssec:asymptotic}

As we have said, quantum non-local operators 
were shown to be related to local charges in  \cite{Bazhanov:1994ft,Bazhanov:1996dr,Bazhanov:1996aq}. 
The first two papers considered critical bulk theories and chiral perturbations and the third paper considered the addition of bulk perturbations. In the first case, they showed that the ``quantum transfer matrices'' they considered had an expansion for small perturbation parameter $\lambda$ as non-local charges, and for large $\lambda$ an asymptotic expansion in terms of local charges. In our language, the quantum transfer matrix
corresponds to a chirally perturbed topological defect $\cD(\lambda \psi)$ in the CFT which in the UV has an expansion in $\lambda$ of integrals along the defect, but for large $\lambda$, the same perturbed defect has an expansion in local charges. For this to be valid, the IR limit of the perturbed defect must be the identity defect. 

We have found models in which we propose the existence of non-local conserved charges as perturbed defects but for which we have not (as yet) found any local conserved currents. 
While we have paid most attention to bulk-perturbed theories, we can choose to turn off the bulk perturbation and then consider the resulting chirally perturbed defects in the unperturbed CFT. Any flow in the IR to an expansion over local conserved charges in the bulk-perturbed model ought then imply a flow of the chirally perturbed defects to an expansion over corresponding local conserved charges in the unperturbed CFT. Such a relation will only be possible if the ultimate IR endpoint of the chirally perturbed defect is the identity defect.

All our calculations so far in this paper have been perturbative, but there is a non-perturbative technique which can be used to investigate the IR endpoint of perturbed defect, which is the truncated conformal space approach (TCSA). This is a numerical approximation to a perturbed theory in finite-dimensional spaces which has already been applied to defect models,
in particular it was used to investigate the joint bulk- and defect-perturbations of the Lee-Yang model in \cite{Bajnok:2013waa}.
In this section we apply this technique to study  chiral defect perturbations in CFT and investigate their IR endpoints. For technical reasons, it is simplest to consider the original system on a cylinder with boundary conditions on the two ends so that in the crossed channel the system is propagating along a strip with boundary conditions on the two sides.

\begin{figure}[t]
\centering
    \includegraphics[width = \linewidth]{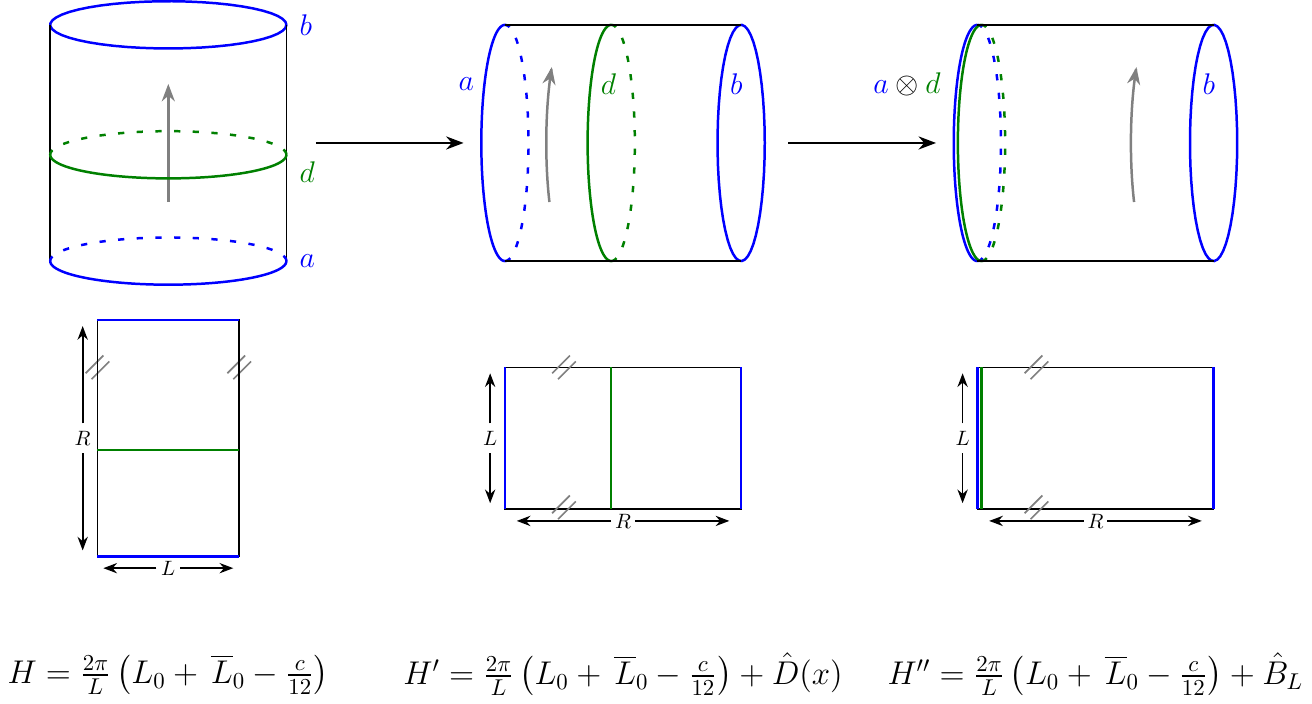}
\caption{The Hamiltonian $H'$ in the crossed channel corresponding to propagation along the defect $d$  on a system on a cylinder of length $R$ with boundary conditions $a$ and $b$ on the ends of the cylinder, and $H''$ for the same system where the defect has been fused to the left boundary. 
}
    \label{fig:2}
\end{figure}

The principal idea is that the nature of an IR fixed point can be constrained or even identified by the spectrum of the corresponding Hamiltonian. While we have, up to now, thought of defects as operators which may or may not commute with a bulk Hamiltonian, we can also look at a crossed channel in which the defect runs along the ``timelike'' direction at a fixed spatial position $x$ and the Hamiltonian for the system with such a defect is the conformal Hamiltonian perturbed by a defect operator $\hat D(x)$ located on the defect, as in Figure~\ref{fig:2}.

Since the chirally-perturbed defect is translationally invariant, it can be moved to the boundary so that spectrum of the Hamiltonian $H'$ on the strip with boundary conditions $a$ and $b$ and a defect $d$ is equal to the spectrum of the Hamiltonian $H''$ of the system on the strip without a defect but with boundary conditions $a\otimes d$ and $b$ perturbed by a boundary field $\hat B_L$. 
For the Lee-Yang model, boundary perturbations were studied extensively in \cite{Dorey:1997yg} (see e.g.\ Figures 1a and 1b there). 

We have now investigated several models with commuting non-local charges.
In each case, we chose the boundary conditions $a$ and $b$ 
to be the identity representation. With no perturbation turned on, the spectrum of the strip with identity boundary conditions on the two edge and a defect of type $d$ is given by the $d$ representation. If the defect flows to the identity in the IR, the IR end point will simply be the system on a strip with identity boundary conditions and its spectrum will be the identity representation; if the spectrum is not given by that of the identity representation, the perturbed defect does not flow to the identity defect in the IR.

In each case we looked at where there are associated local charges, the spectrum of the perturbed defect flows in the IR to the identity, and in each case where there are no associated local charges, the defect does not flow to the identity defect in the IR. 

\begin{figure}[t!]
    \centering
    \begin{subfigure}{0.3\textwidth}
      \includegraphics[width=\textwidth]{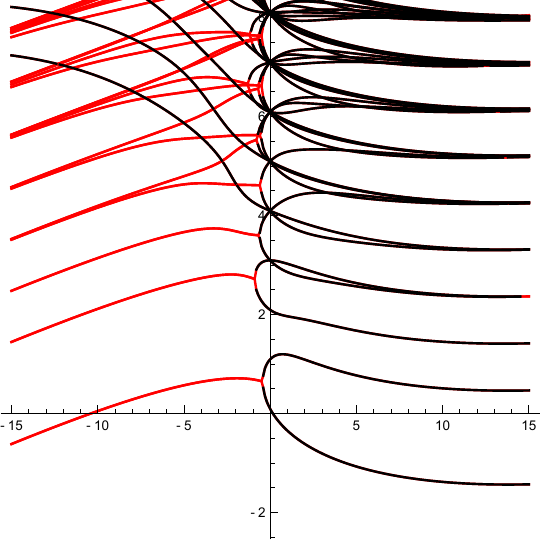}
      \caption{ $D_{(1,2)} + \lphi_{(1,3)}$}
      \label{fig:310defectflowsA}
    \end{subfigure}
    \begin{subfigure}{0.3\textwidth}
      \includegraphics[width=\textwidth]{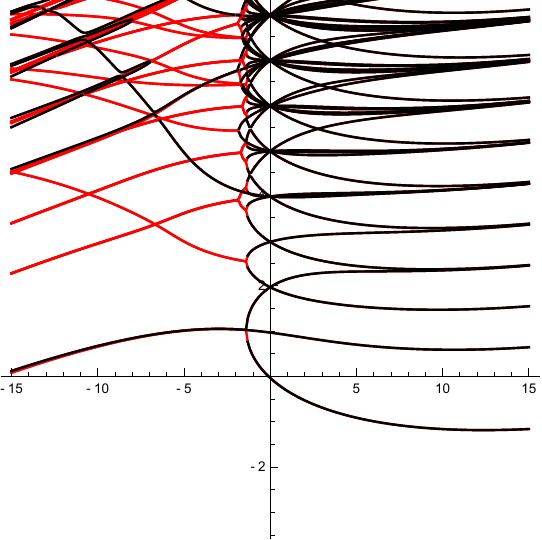}
      \caption{ $D_{(1,3)} + \lphi_{(1,5)}$}
      \label{fig:310defectflowsB}
    \end{subfigure}
        \begin{subfigure}{0.3\textwidth}
      \includegraphics[width=\textwidth]{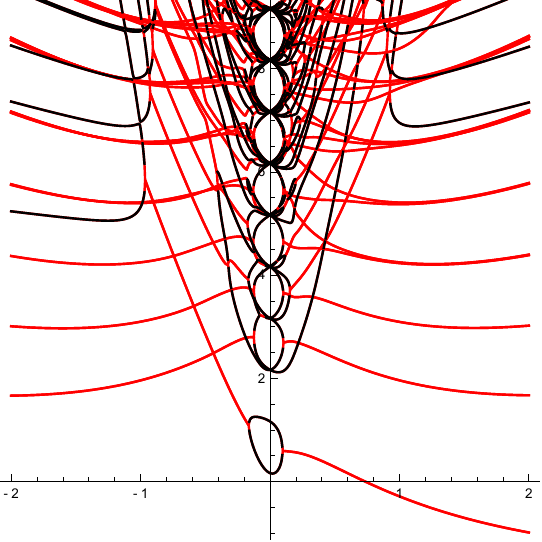}
      \caption{ $D_{(1,5)} + \lphi_{(1,7)}$}
      \label{fig:310defectflowsC}
    \end{subfigure}\\\medskip
        \centering
    \begin{subfigure}{0.3\textwidth}
      \includegraphics[width=\textwidth]{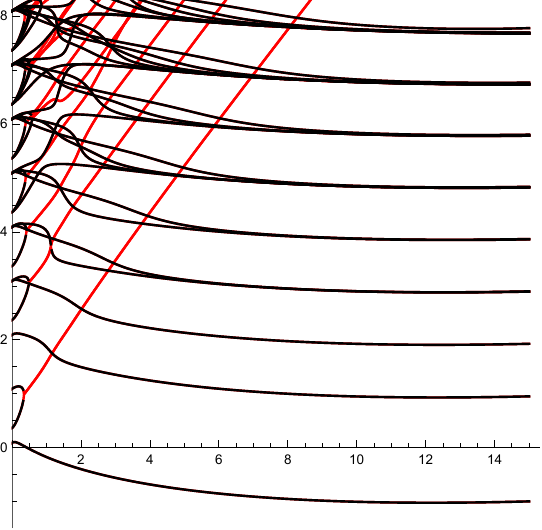}
      \caption{ $D_{(1,1)} \oplus D_{(1,2)}+ \lphi_{(1,2)}$}
      \label{fig:310defectflowsD}
    \end{subfigure}
    \begin{subfigure}{0.3\textwidth}
      \includegraphics[width=\textwidth]{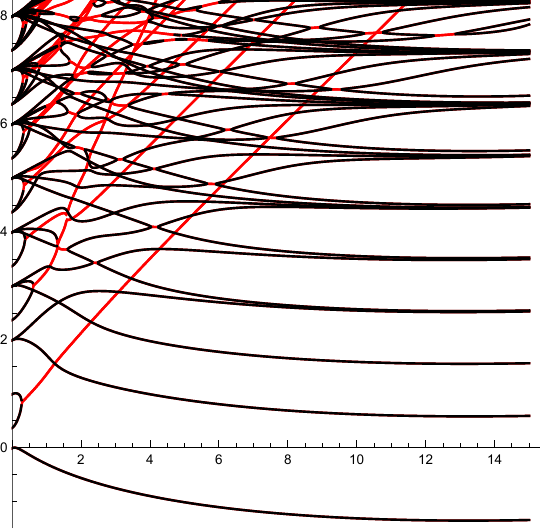}
      \caption{ $D_{(1,1)} \oplus D_{(1,4)} + \lphi_{(1,4)}$}
      \label{fig:310defectflowsE}
    \end{subfigure}
        \begin{subfigure}{0.3\textwidth}
      \includegraphics[width=\textwidth]{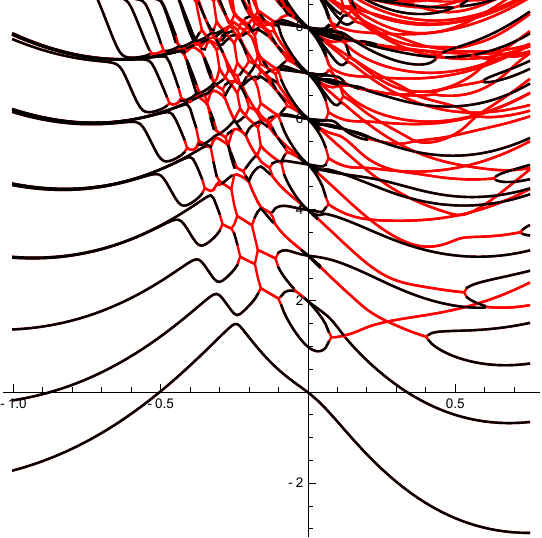}
      \caption{ $D_{(1,4)} + \lphi_{(1,7)}$}
      \label{fig:310defectflowsF}
    \end{subfigure}
    \caption{The lowest 77 levels in the spectrum of the perturbed defect Hamiltonian on a strip in the model $M(3,10)$ for various perturbations as calculated in TCSA with truncation up to level 18. The black lines correspond to real eigenvalues, the red lines to eigenvalues with non-zero imaginary part. In all cases a function of the perturbation parameter has been added to the eigenvalues for reasons of clarity.
    Plots (a), (b) and (d) have clear line crossing of black lines (an indication of integrability), while in (e) and (f) there are clear line avoidances; plot (c) is more ambiguous.  
    Note that in cases (d) and (e) we only show the graphs for positive values since the perturbation is $\mathbb{Z}_2$ symmetric.}
    \label{fig:310defectflows}
\end{figure}

We illustrate this in the case of the model $M(3,10)$ in Figure~\ref{fig:310defectflows}.

We show plots of the spectrum of the purely chiral defect Hamiltonian against the perturbation parameter, calculated using the TCSA method used in \cite{Dorey:1997yg}, for three classes of models. In these plots, black lines indicate real energies, red lines indicate energies with non-zero imaginary part. In each case there is a non-trivial ground state energy which we have adjusted for clarity - the difference in behaviour between the perturbations by $(1,2),(1,3),(1,4),(1,5)$ on the one hand and $(1,7)$ on the other is that the first four perturbations have $h<1/2$ and a finite ground state energy while the $(1,7)$ perturbation has $h>1/2$ and so the ground state energy diverges as the truncation level increases.

It is also often the case that one can see strong indications of integrability, or lack of integrability, of the perturbed Hamiltonian from the presence, or absence, of line-crossings. It is common folklore that the spectrum can only have degeneracies if there is some conserved quantity which distinguishes states of the same energy. In TCSA, the only exact symmetries are discrete symmetries; those from local conserved currents being broken explicitly by the truncation, so one can only have an indication of integrability, not a proof.

We identified each perturbed defect flow as falling into one of the following three classes:
\begin{itemize}
    \item[i)] The massive perturbation has local conserved charges and we find the commuting non-local charges as a perturbed defect. 
    
    This is the case for the perturbations $D_{(1,1)}\oplus D_{(1,2)}+\lambda\psi_{(1,2)}$ (Figure \ref{fig:310defectflowsD}), $D_{(1,2)}+\lphi_{(1,3)}$  (Figure \ref{fig:310defectflowsA}) and $D_{(1,3)}+\lphi_{(1,5)}$ (Figure \ref{fig:310defectflowsB}). 
    In each case
    there is a flow in which the IR spectrum is that of the identity representation (indicating the IR endpoint is the identity defect).

    The plots also exhibit multiple line crossings (an indication that the perturbation is integrable)
\item[ii)] The massive perturbation has no local conserved charges and have not found any non-local charges as a perturbed defect. 

This is the case for the perturbations $D_{(1,1)}\oplus D_{(1,4)}+\lambda\psi_{(1,4)}$ (Figure \ref{fig:310defectflowsE}), $D_{(1,4)}+\lambda\psi_{(1,7)}$ (Figure \ref{fig:310defectflowsF}). 
In each case 
there is a flow in which the IR spectrum is that of the identity representation (indicating the IR endpoint is the identity defect).

There are also multiple line avoidances, suggesting that the perturbation is not integrable.

    \item[iii)] The massive perturbation has no local conserved charges, but there are a family of commuting non-local charges. 
    
    This is the case for $D_{(1,5)}+\lambda\psi_{(1,7)}$ (Figure \ref{fig:310defectflowsC}). In this case, in neither direction of the flow does the spectrum flow to that of the vacuum representation - in both directions, the ground states are a pair with complex conjugate energies, so in neither direction does the defect flow to the identity defect. 
    
    The evidence on line crossings is more ambiguous - there are multiple possible line crossings indicating that the perturbation may well be integrable, but there are also large renormalisation effects due to the divergence of the perturbation which would require much more work to reduce.
    \end{itemize}

To summarise,
for the known integrable perturbations, (1,2), (1,3) and (1,5), in each case we find that there is a flow to the identity defect. These all exhibit the ``typical'' characteristic of integrable theories of line-crossings from the UV to the IR, and the flow to the IR to a single identity defect implies that in this limit the system can be described by a perturbation by local conserved currents.

For the perturbations by (1,4) and (1,7) we considered, we find that only the case $D(1,5)+\lphi_{1,7}$ exhibits possible line-crossings, but that there is no IR limit in which the system flows to a single identity defect. The presence of line-crossings suggests the flow may be integrable, but we cannot deduce the existence of local conserved quantities since there is no flow to the identity defect. This is consistent with the results of our search in section \ref{sssec:17perts} which did not reveal any local currents with spins less than 20.

\section{Outlook}\label{sec:outlook}
Maybe the most important conceptual insight of this paper is that as a next step in the investigation of symmetries in QFT one should try to go from topological symmetries to rigidly translation invariant symmetries. One crucial property of the latter is that they still allow for a non-singular fusion procedure. Conformal defects, on the other hand, in general have singular fusion, see e.g.\ \cite{Bachas:2007td,
Bachas:2013ora,
Konechny:2015qla,
Diatlyk:2024zkk}.

We have introduced the bulk commutation relation in SymTFT and chiral TFT language as a tool to investigate such rigidly translation invariant defects in perturbed two-dimensional conformal field theories from the point of view of the unperturbed theory. In the study of minimal models, we found that including these defects can greatly enhance the symmetry at one's disposal in perturbed theories, as compared to just asking for defects that remain fully topological in the perturbed model.

\medskip

There are many open questions and directions for further research building on the results we presented here. We hope to return to some of these points in future work. 

The only perturbation not of the form $(1,2)$, $(1,3)$, or $(1,5)$ that we found which does not require regularisation is the $(1,4)=(1,7)$ perturbation of $M(2,11)$. An important question is therefore to understand if there is a regularisation procedure which is compatible with the bulk commutation condition in the sense that the perturbed defect continues to commute with the perturbed Hamiltonian. This would be particularly interesting for the $(1,7)$-perturbations in $M(3,10)$ and $M(5,18)$ in order to compare them to the flows investigated in \cite{Klebanov:2022syt,Delouche:2024tjf,Nakayama:2024msv,Katsevich:2024jgq,Katsevich:2024sov,Ambrosino:2025xsv}.\footnote{
In \cite{Nakayama:2024msv,Ambrosino:2025xsv} an infinite number of flows is considered, namely $M(p,pk+I) \to M(p,pk-I)$ induced by a $(1,2k+1)$-perturbation. Apart from $k=1,2$, we have (so far) only found a solution to the bulk commutation condition 
in the case $k=3$ with $(p,I) = (2,3)$, $(2,5)$, $(3,1)$, $(5,3)$;
for $k=4$ with $(p,I) = (2,3)$, $(3,2)$; and
for $k=5$ and $k=6$ with $(p,I) = (2,3)$.
}

The formalism we present also makes it possible to find the functional relations between different non-local conserved charges. For the $(1,3)$-perturbation this has been analysed in \cite{Runkel:2010ym}. In terms of the formulation as Yetter-Drinfeld modules briefly mentioned in Section~\ref{sec:rigid-trans}, this amounts to finding instances where the tensor product of two modules forms a (possibly non-split) exact sequence. One should verify that also in the case of $(1,2)$ and $(1,5)$ one recovers the expected T-system functional relations \cite{Kuniba:1993cn,Fioravanti:1995cq,Bazhanov:2001xm}. Much more interesting of course is what happens in the new examples, such as $(1,7)$. This again relies on a regularisation procedure, but the hope would be that one can pass through the integrable tool chain from T-system to Y-system to non-linear integral equations. These could then be compared to the proposed non-linear integral equations in \cite{Ambrosino:2025xsv}. If that indeed works, this would be an indication that the approach via non-local conserved charges is a more fundamental criterion for integrability than the existence of local conserved charges as the latter are absent for example in the $(1,7)$-perturbation of $M(5,18)$.

The existence of mutually commuting families of conserved charges is not, however, in itself sufficient to prove integrability. There is further work to be done to show that standard quantities such as the spectrum can be found using these new defects.

The rigidly translation invariant defects do not only exist on cylinders with periodic boundary conditions, but one can also investigate them on a strip (possibly with suitably perturbed boundary conditions), or on a cylinder with twisted boundary conditions. For the latter case it is for example possible to twist by another rigidly translation invariant defect, and to use a solution for the exchange matrix $M$ (Section~\ref{sec:commute-via-M-matrix}) at the crossing point with the defect wrapping the cylinder. This indicates that there are interesting generalisations of the strip and tube algebras, whose purely topological variants have for example been used to analyse the field content of various models, see e.g.\ \cite{Chang:2018iay,Konechny:2019wff,Lin:2022dhv,Bartsch:2023wvv,Bhardwaj:2023idu,Cordova:2024vsq}. 

A second reason to investigate perturbed defects in the crossed channel is that the spectrum of the Hamiltonian on the twisted cylinder is amenable to the truncated conformal space approach and this can serve as an important indicator of integrability (see \cite{Bajnok:2013waa} and Section~\ref{ssec:asymptotic}). It would be interesting to apply this method to the new perturbations we found in this paper.

Furthermore, we have only investigated in detail certain classes of perturbed defects -- these are defects which are perturbed by a single fixed field and which commute with the corresponding bulk perturbation. 
In the unperturbed CFT, all choices of (holomorphic or antiholomorphic) perturbing defect field result in rigidly translation invariant defects, and it would be interesting to investigate their interplay, e.g.\ which perturbing fields lead to commuting families, and 
whether perturbed defects for different perturbing fields commute or not.

All of our explicit expressions and examples focussed on diagonal conformal field theories: we only considered the bulk commutation condition in detail for chiral TFT with trivial surface defect, and we only searched for solutions in A-type Virasoro minimal models. An important next step is therefore to investigate the bulk commutation condition for non-trivial surface defects as in Figure~\ref{fig:geometry}. One can then ask how solutions behave under gauging topological symmetries (aka taking generalised orbifolds \cite{Frohlich:2009gb}). We saw the need for this already in the $M(3,10)$ model studied in Section~\ref{sec:M310}. The case where one passes between spin theories and oriented theories seems particularly interesting as spin theories can also have local conserved currents of half-integer weight. The relevant formalism on the CFT side to do this is already in place \cite{Runkel:2020zgg,Hsieh:2020uwb,Runkel:2022fzi}.

On the more speculative side, one can ask what happens in higher dimensional QFTs. For example, one could try perturb a $q$-dimensional topological defect by suitable $p$-dimensional operators for $p<q$ in a $d$-dimensional QFT. 

\paragraph{Acknowledgements}
We thank Davide Gaiotto, Marius de Leeuw, Arshia Moghanjoghi, and Yifan Wang for discussions.
GW thanks the Fachbereich Mathematik in Hamburg for hospitality when this research was started.
The authors thank  the  Deutsche Forschungs\-gemeinschaft (DFG, German Research Foundation) under Germany's Excellence Strategy - EXC 2121 ``Quantum Universe'' - 390833306,  and the Collaborative Research Center - SFB 1624 ``Higher structures, moduli spaces and integrability'' - 506632645, for support. GW was, in addition, supported by the STFC under grant ST/T000759/1. 
For the purposes of open access, the authors have applied a Creative Commons Attribution (CC BY) license to any Accepted Author Manuscript version arising from this submission.

\appendix

\section*{Appendices}

\section{F-matrix in the case \boldmath \texorpdfstring{$x \otimes x \cong \mathbf{1}$}{x . x = 1}}\label{app:F-matrix-xx=1}

Let $x \in \mathrm{Irr}(\mathcal{F})$ satisfy $x \otimes x \cong \mathbf{1}$, and let $a \in \mathrm{Irr}(\mathcal{F})$ be a fixed point for $x$, that is, $x \otimes a \cong a$. Note that $x \otimes x \cong \mathbf{1}$ implies $x = \ov x$. We are interested in the F-matrix element $\F{xax}{a}{aa}$. It is given by evaluating
\begin{equation}\label{eq:xx=1-aux1}
\F{xax}{a}{aa} \quad \includegraphics[valign = c]{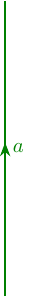}\quad  = \quad \includegraphics[valign = c]{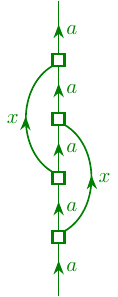} \quad = \quad \includegraphics[valign = c]{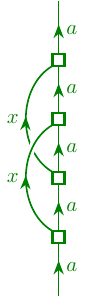}  
\end{equation}
where in the second step we used the R-matrix (Figure~\ref{fig:TFT-line-rules}) to rotate the $x$-loop from the right to the left. Now substitute
\begin{equation}
\includegraphics[valign = c]{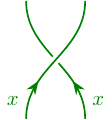} \quad = \quad  \R{xx}{1} \quad  \includegraphics[valign = c]{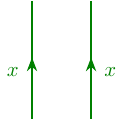}
\end{equation}
which follows from the assumption $x \otimes x \cong \mathbf{1}$ by composing with the preferred basis element $x \otimes x \to \mathbf{1}$ from the top. 

After the substitution, the rightmost diagram in \eqref{eq:xx=1-aux1} turns into two basis / dual basis pairs, and we arrive at
\begin{equation}
	\F{xax}{a}{aa} = \R{xx}{1} \ .
\end{equation}
The possible values of $\R{xx}{1}$ for a simple object of order 2 are $\{ \pm 1, \pm i\}$. But we could have chosen to rotate the loop above the other way, in which case we would have obtained $\F{xax}{a}{aa} = 1/\R{xx}{1}$. This is only in agreement with the first computation if $\R{xx}{1} \in \{ \pm 1\}$.

\section{Bulk commutation condition and mutual commutativity for \boldmath \texorpdfstring{$\mathcal{D} = \mathbf{1} \oplus x$}{D = 1 + x}}\label{app:2-channel-fusion-proofs}

Here we consider the bulk commutation condition and mutually commutativity in the case $\mathcal{D} = \mathbf{1} \oplus x$, where $x$ is the perturbing field. We assume that $\mathcal{F}$ is a ribbon fusion category, and that $x \in \mathcal{F}$ is simple and satisfies $x \otimes x \cong \mathbf{1} \oplus y$, where $y \not\cong \mathbf{1}$ and $y$ is simple (hence in particular non-zero). 

\subsection{F- and R-matrices}\label{app:FR-xx=1+y}

Since $\mathbf{1}$ appears in the tensor product, $x$ is self-dual, $x = \ov x$. For later use we note
\be 
\includegraphics[valign = c]{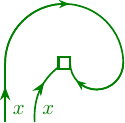}\; = \; \nu_x \raisebox{-0.5cm}{\includegraphics[valign = c]{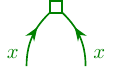}}\comma \quad \raisebox{-.28cm}{\includegraphics[valign = c]{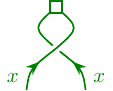}} = \;\includegraphics[valign = c]{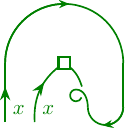} \;= \nu_x \,\theta^{-1}_x \raisebox{-0.5cm}{\includegraphics[valign = c]{Figs/appAfig1b.pdf} }
\ee
where $\nu_x$ is the Frobenius-Schur indicator of the self-dual object $x$ as already encountered in \eqref{eq:Frob-Schur-dim}. This implies
\be\label{eq:R-nu-theta}
    \R{xx}{1} = \nu_x \theta_x^{-1} \ .
\ee
We abbreviate 
\be\label{eq:1+x-solution-abbrev}
d = \dim(x) 
~,~~
\theta = \theta_x
~,~~
\nu = \nu_x
~,~~
\alpha = \frac{d\theta-\nu \theta^{-1}}{d^2-1}
~,~~
\beta = \frac{d\theta^{-1}-\nu \theta}{d^2-1} \ .
\ee
By assumption, $\mathrm{Hom}(x \otimes x, x \otimes x)$ is two-dimensional and therefore any three morphisms $x \otimes x \to x \otimes x$ must satisfy a linear relation. We claim:
 \be \begin{split}\label{eq:Kauffman-relation}
  \includegraphics[valign = c]{Figs/appfig2a.pdf} = \quad \alpha \; \includegraphics[valign = c]{Figs/appfig2b.pdf} + d \,\nu \,\beta  \includegraphics[valign = c]{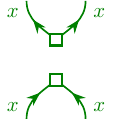} \\\\
    \includegraphics[valign = c]{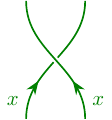} =\quad \beta\;\includegraphics[valign = c]{Figs/appfig2b.pdf} + d \,\nu \,\alpha    \includegraphics[valign = c]{Figs/appkauf.pdf}
\end{split}
 \ee
To check this, insert both sides of each equation into the following diagrams:
\be
\includegraphics[valign = c]{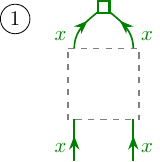}\qquad \qquad 
\includegraphics[valign = c]{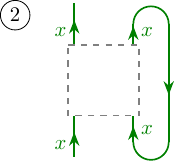}\qquad 
\ee
For example, the first equation in \eqref{eq:Kauffman-relation}  produces
\be
    \circled{1} ~:~ \nu\theta^{-1} = \alpha + d \nu \beta 
    \quad , \qquad
    \circled{2} ~:~ \theta = \alpha \, d + \nu \, \beta \ ,
\ee
where in $\circled{1}$ we used \eqref{eq:R-nu-theta}. It is easy to check that these indeed hold for the values of $\alpha$ and $\beta$ as above. We also note that composing both equations in \eqref{eq:Kauffman-relation} with the chosen basis element in $x \otimes x \to y$ yields
\be
    \R{xx}y = \alpha = \beta^{-1} \ .
\ee
The equality $\alpha=\beta^{-1}$ can be equivalently written as
\be\label{eq:xx=1+y_condition_on_data}
    (d\theta-\nu \theta^{-1})(d\theta^{-1}-\nu \theta)
    =(d^2-1)^2 
    \quad \overset{d \neq 0}\Leftrightarrow \quad
    d^3 - 3d + \nu (\theta^{-2} + \theta^2) = 0
\ee
This relation between quantum dimension, Frobenius-Schur indicator, and ribbon twist (all of $x$) is implied by the fusion rule $x \otimes x \cong \mathbf{1} \oplus y$ that we assumed at the outset.

\medskip

We will also need the F-matrix $\F{xxx}{x}{ab}$ with $a,b \in \{\mathbf{1},y\}$, as well as $\R{xy}x$ and $\R{yx}x$. To this end, fix basis elements in $x \otimes y \to x$ and $y \otimes x \to x$ as follows (these Hom-spaces are one-dimensional as $x \otimes x \cong 1 \oplus y$):
\be 
\includegraphics[valign = c]{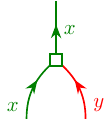} = \; \zeta \; \includegraphics[valign = c]{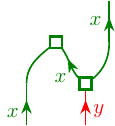}\qquad \qquad \includegraphics[valign = c]{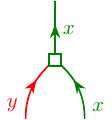} = \; \xi \; \, \includegraphics[valign = c]{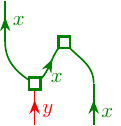}
\ee
Here $\zeta,\xi \in \mathbb{C}^\times$ are constants which parametrise different choices of basis -- the final results have to be independent of these. 
The dual basis is (in the sense that they compose to $\mathrm{id}_x$ with the corresponding element above) 
\be 
\includegraphics[valign = c]{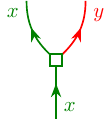} = \; \frac{1}{\zeta} \frac{1}{1-d^2} \; \includegraphics[valign = c]{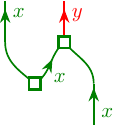}\qquad \qquad \includegraphics[valign = c]{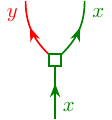} = \; \frac{1}{\xi} \frac{1}{1-d^2} \; \includegraphics[valign = c]{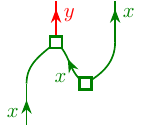}
\ee
The projector onto the summand $y$ in $x \otimes x \cong \mathbf{1} \oplus y$ can be expressed as
\be 
\includegraphics[valign = c]{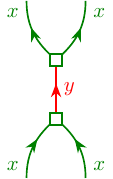}\; = \; \includegraphics[valign = c]{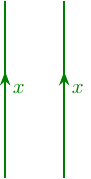}  \; \; - \; \includegraphics[valign = c]{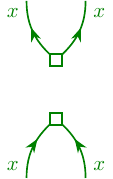}
\ee
Combining these ingredients, it is now straightforward to compute the R- and F-matrix by evaluating the three diagrams 
\be 
\includegraphics[valign = c]{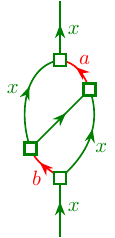} =  \F{xxx}{x}{ab} \;\;\includegraphics[valign = c]{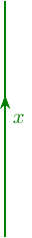}  \qquad\quad\; \includegraphics[valign = c]{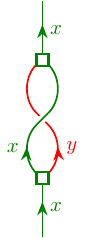}  =  \R{xy}{x} \; \;\includegraphics[valign = c]{Figs/bubbleb.pdf}\qquad\quad\;  \includegraphics[valign = c]{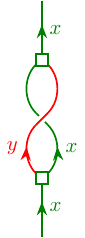} =  \R{yx}{x} \; \;\includegraphics[valign = c]{Figs/bubbleb.pdf}
\ee
We find (here $\mathrm{F}_{ab} = \F{xxx}{x}{ab}$, $\mathrm{G}_{ab} = \G{xxx}{x}{ab}$)
\be
   \begin{pmatrix}
\mathrm{F}_{11} & \mathrm{F}_{1y} \\
\mathrm{F}_{y1} & \mathrm{F}_{yy} 
    \end{pmatrix}
=    
   \begin{pmatrix}
\frac{\nu}{d} &  \zeta \,(1-d^{-2}) \\
\xi^{-1} & - \frac{\zeta}{\xi}  \, \frac{\nu}{d}
    \end{pmatrix}
~~ , \quad
    \begin{pmatrix}
\mathrm{G}_{11} &  \mathrm{G}_{1y} \\
\mathrm{G}_{y1} &  \mathrm{G}_{yy} 
    \end{pmatrix}
=    
   \begin{pmatrix}
\frac{\nu}{d} &  \xi \,(1-d^{-2}) \\
\zeta^{-1} & - \frac{\xi}{\zeta}  \, \frac{\nu}{d}
    \end{pmatrix} 
\ee
and 
\be
\R{xy}x = \frac{\xi}{\zeta}\,\beta \, \nu \, \theta^{-1}
\quad , \quad
\R{yx}x = \frac{\zeta}{\xi}\,\beta \, \nu \, \theta^{-1} \ .
\ee

We conclude this preparatory section by giving the values for $d,\nu,\theta,\zeta,\xi$ in the example of the minimal model $M(p,q)$ with $x = (1,2)$, $y = (1,3)$ (recall that $t=p/q$):
\be\begin{split}
    &d = \frac{S_{1x}}{S_{11}} = -\frac{\sin(2 \pi t)}{\sin(\pi t)}
    ~,~~
    \nu = +1
    ~,~~
    \theta = e^{2 \pi i h_x} = e^{\pi i (3 t-2)/2} \ ,
    \\
    &\zeta = \xi = -2 \cos(\pi t)\,\frac{ \Gamma(2 - 2 t) \Gamma(-1 + 3 t)}{\Gamma(1 - t) \Gamma(2 t)} \ .
\end{split}\ee
The solution we present here also applies to $x = (1,q-2)$, $y=(1,3)$, with appropriately adapted constants.

\subsection{Solving the bulk commutation condition}

We can now complete the verification of the bulk commutation condition from Section~\ref{sec:xx=1+y-solution}.
The second identity in \eqref{eq:1+x-comm-cond} becomes
\begin{equation}
    \frac{\R{xx}{1} - (\R{xx}{1})^{-1} }{(\F{xxx}{x}{11})^{-1}-1}
    = \frac{1}{\R{xx}{y}} - \R{xx}{y} 
    \quad \Leftrightarrow \quad
    \frac{\theta^{-1} - \theta }{d-\nu}
    = \beta - \alpha \ . 
\end{equation}
It is easy to check that the second equality does indeed hold for the values of $\alpha$ and $\beta$ given in \eqref{eq:1+x-solution-abbrev}.

\subsection{Perturbed defects mutually commute}

Next we check that the perturbed defects mutually commute, as claimed in Section~\ref{sec:xx=1+y-solution}.
Namely, we solve the condition in Section~\ref{sec:commute-via-M-matrix} under the assumptions made in the beginning of Appendix~\ref{app:2-channel-fusion-proofs}. We take $\mathcal{D} = \mathcal{D}' = \mathbf{1} \oplus x$, but we keep different choices of defect field coefficients (see \eqref{eq:psi-barpsi-expansion}) $\kappa_{ab}$, $\kappa'_{ab}$, $\wt\kappa_{ab}$, $\wt\kappa'_{ab}$ on the two defects. Recall from \eqref{eq:D=1x-lamxx=0} that we take $\kappa_{xx} = \wt\kappa_{xx} = \kappa'_{xx} = \wt\kappa'_{xx} = 0$ and so the non-zero defect field coefficients are
\begin{equation}
    \kappa_{1x} ~~,\quad \kappa_{x1} ~~,\quad \kappa'_{1x} ~~,\quad \kappa'_{x1} \ ,
\end{equation}
and analogously for $\wt\kappa_{ab}$. The coefficients $\kappa_{ab}$ and $\wt\kappa_{ab}$ are related by \eqref{eq:1+x-comm-cond},
\be
\wt\kappa_{x1} = \frac{\Lambda}{\kappa_{1x}} ~,~~
\wt\kappa_{1x} = \frac{\Lambda}{\kappa_{x1}} 
\quad \text{where}~~
\Lambda = \frac{1}{\R{xx}{y}} - \R{xx}{y} = \beta-\alpha \ ,
\ee
The same relations hold for $\kappa'_{ab}$ and $\wt\kappa'_{ab}$.

There are nine entries in $M_{ab}^{cd}(m)$ in \eqref{eq:ansatz-M} which can be non-zero, and we arrange them into matrices according to the label $m$ of the summand on the fused defect:
\begin{equation}\label{eq:M-into-matrices}
\begin{pmatrix}
    M^{11}_{11}(1) & M^{11}_{xx}(1)
    \\
    M^{xx}_{11}(1) & M^{xx}_{xx}(1)
\end{pmatrix}
~~ , \quad
\begin{pmatrix}
    M^{1x}_{1x}(x) & M^{1x}_{x1}(x)
    \\
    M^{x1}_{1x}(x) & M^{x1}_{x1}(x)
\end{pmatrix}
~~ , \quad
M^{xx}_{xx}(y) \ .
\end{equation}
We can now write the conditions in \eqref{eq:M-matrix-cond1} and \eqref{eq:M-matrix-cond2} for these variables, where $a,b,c,d \in \{ \mathbf{1},x \}$, $m,u \in \{\mathbf{1},x,y\}$. For example, for $a=b=c=u=1$ and $d=m=x$ we get the conditions
\be\begin{split}
\kappa_{1x} \, M_{x1}^{1x}(x)
 +
 \kappa'_{1x} \, M_{1x}^{1x}(x)
 &~=~ 
  \kappa'_{x1} \, M_{11}^{xx}(1) \, \mathrm{G}^{(xxx)x}_{11}
 +
 \kappa_{1x} \, M_{11}^{11}(1) \ ,
 \\
\frac{\Lambda}{\kappa_{x1}} \, M_{x1}^{1x}(x)
+
\frac{\Lambda}{\kappa'_{x1}} \, M_{1x}^{1x}(x)
&~=~
\frac{\Lambda}{\kappa'_{1x}} \, M_{11}^{xx}(1) \, \frac{1}{\R{xx}{1}} \, \F{xxx}{x}{11}
+
\frac{\Lambda}{\kappa_{x1}}  \, M_{11}^{11}(1) \ .
\end{split}\ee
Altogether, these conditions form a homogeneous system of 24 linear equations for 9 variables. It turns out that there is a one-dimensional space of solutions, which is spanned by:
{\allowdisplaybreaks
\begin{align}
M^{11}_{11}(1) &= 
\frac{(d - \nu) \theta^2  (\kappa_{1x} \kappa_{x1})^2 + \big(\nu - d \nu \theta + \theta^2 (-d + \theta)\big) 
\kappa_{1x} \kappa_{x1}  \kappa'_{1x}  \kappa'_{x1} 
+ (d \nu - 1) \theta  (\kappa'_{1x} \kappa'_{x1})^2}{(\theta^2 - 1) (\theta  \kappa_{1x}  \kappa_{x1} - \nu  \kappa'_{1x}  \kappa'_{x1})}
\nonumber \\    
M^{11}_{xx}(1) &= 
\frac{
d  \kappa_{1x}  \kappa'_{1x} ( \kappa_{1x}  \kappa_{x1} -  \kappa'_{1x}  \kappa'_{x1})}{
\theta  \kappa_{1x}  \kappa_{x1} - \nu  \kappa'_{1x}  \kappa'_{x1}}
\nonumber \\    
M^{xx}_{11}(1) &=  
\frac{
\theta  \kappa_{x1}  \kappa'_{x1} ( \kappa_{1x}  \kappa_{x1} -  \kappa'_{1x}  \kappa'_{x1})}{
\theta  \kappa_{1x}  \kappa_{x1} - \nu  \kappa'_{1x}  \kappa'_{x1}}
\nonumber \\    
M^{xx}_{xx}(1) &= 
\frac{
(d \nu - 1) \theta  (\kappa_{1x} \kappa_{x1})^2 + \big(\nu - d \nu \theta + \theta^2 (-d + \theta)\big)  \kappa_{1x}  \kappa_{x1}  \kappa'_{1x}  \kappa'_{x1} + (d - \nu) \theta^2  (\kappa'_{1x}  \kappa'_{x1})^2}{(\theta^2 - 1) (\theta  \kappa_{1x}  \kappa_{x1} - \nu  \kappa'_{1x}  \kappa'_{x1})}
\nonumber \\    
M^{1x}_{1x}(x) 
&= \kappa_{1x} \kappa'_{x1} \ ,
\nonumber \\    
M^{1x}_{x1}(x)
&= M^{x1}_{1x}(x) = \frac{(d - \nu) \, \theta \, (\kappa_{1x} \kappa_{x1} - \kappa'_{1x} \kappa'_{x1})}{\theta^2 - 1} \ ,
\nonumber \\    
M^{x1}_{x1}(x)
&=
\kappa_{x1} \kappa'_{1x} \ ,
\nonumber \\    
M^{xx}_{xx}(y) 
&= \frac{\big(1 + d^2 \theta^2 - d \nu (1 + \theta^2)\big)\, \kappa_{1x} \kappa_{x1}  - 
  \big(d^2 + \theta^2 - d \nu (1 + \theta^2)\big) \, \kappa'_{1x} \kappa'_{x1}
  }{(d^2-1) (\theta^2 - 1)} \ .
\end{align}
}In order to verify that the above coefficients indeed solve \eqref{eq:M-matrix-cond1} and \eqref{eq:M-matrix-cond2} one needs to make use of the relation in \eqref{eq:xx=1+y_condition_on_data}.

For special values of $\kappa_{1x} \kappa_{x1} / \kappa'_{1x}  \kappa'_{x1}$, the resulting $M$ in \eqref{eq:ansatz-M} is not invertible or not well-defined, but generically it is. Invertibility amounts to all three matrices in \eqref{eq:M-into-matrices} being simultaneously invertible. At the special values, we expect the defect operators to obey $T$-system type functional relations. Related functional equations were investigated in \cite{Runkel:2007wd,Manolopoulos:2009np,Runkel:2010ym,Buecher:2012ma}.

Altogether, away from values of $\zeta := \kappa_{1x} \kappa_{x1} / \kappa'_{1x}  \kappa'_{x1}$ where the matrices \eqref{eq:M-into-matrices} are not well-defined or not invertible, the coefficients above make the topological junction $M$ in \eqref{eq:ansatz-M} solve \eqref{eq:M-matrix-cond1} and \eqref{eq:M-matrix-cond2}, and thus the perturbed defects mutually commute away from the singular values of $\zeta$.
The singular values are isolated points in the complex $\zeta$-plane, and since all matrix elements of the commutator vanish away from these points, by holomorphicity in $\zeta$ they will also vanish at these points.

\section{Details of the calculation of a local conserved current}
\label{app:localcharge}

Let us consider a holomorphic field $W(z)$ in the CFT of weight $h_W$. 
According to \cite{Zamolodchikov:1989hfa}, the first order term $V_1$ in \eqref{eq:zampert} can be 
calculated using\footnote{Note that in this appendix we revert to more standard CFT conventions and denote the field $\varphi$ inserted at the point $z$ by $\varphi(z,\overline z)$ to show clearly that it is not a chiral field.}
\begin{align}
 V_1
 = 
  \oint_z \varphi(\zeta,\overline \zeta)\,W(z)\,\frac{d\zeta}{2\pi i}
 \;,
 \label{eq:dworig}
 \end{align}
 where the contour is a small circle around $z$.
 
At first sight, the dependence of the integrand on $\overline\zeta$ suggests that the integral is not a complex integral, but simply a line integral and dependent on the contour. However,  
the operator product can be replaced by its expansion,
\begin{align}
\varphi(\zeta,\overline \zeta)\,W(z)
= \sum_{n=0}^{h_W-1} \frac{1}{(\zeta-z)^{h_W-n}} \varphi^{(n)}(z,\overline z) + O(1)
\;,
\label{eq:phiwope}
\end{align}
(where $\varphi^{(n)}$ is a level $n$ descendant of $\varphi$) and 
hence the dependence of the integrand on $\overline\zeta$ vanishes in the limit that the contour shrinks to a small circle around $z$.
Now applying the residue theorem, we see that the condition that $V_1$ is a total $\partial$-derivative is that the coefficient of the first order pole, $\varphi^{(h_W-1)}$, is a total derivative. 
By expanding the OPE \eqref{eq:phiwope} about $\zeta$, instead of $z$, we have
\begin{align}
    \varphi(\zeta,\overline\zeta)\,W(z)
    = \sum_{n=0}^{h_W-1}
    \frac{1}{(z-\zeta)^{h_W-n}}
     \hat W_{-n}\varphi(\zeta,\overline\zeta)
+O(1)\;,     
\end{align}
and so 
the state corresponding to the field $\varphi^{(h_W-1)}$ can be written
\begin{align}
    \ket{\varphi^{(h_W-1)}}
    = (-1)^{h_W-1} W_{-h_W+1}\ket\varphi +
      L_{-1}\left[...\right]
\;.
\end{align}

It is perfectly possible to calculate explicitly $ W_{-h_W+1}\ket\varphi$ for arbitrary choices of $W$ and test if they are total $L_{-1}$ derivatives, but it is much simpler to follow \cite{Zamolodchikov:1989hfa} and define (in a similar way)
\begin{align}
    D_n = \oint \varphi(z,\overline z) z^{n} \frac{dz}{2\pi i}\;.
\end{align}
Then 
\begin{align}
   \ket{\varphi^{(h_W-1)}} = 
   D_0 \ket{W} \;,
\end{align}
and the action of $D_0$ can be calculated using
\begin{align}
    [L_m,D_n] = ((\Delta_\varphi -1 )(m+1)-n) D_{n+m}
    \;,\;\;
    D_{-n-1}\ket 0 = \frac{1}{n!} L_{-1}^n \ket\varphi\;.
\end{align}
If we can show that 
\begin{align}
    D_0 W_{-h_w}\ket 0  \neq L_{-1} N_{-h_w+1}\ket\varphi\;,
    \label{eq:pcond}
\end{align}
for any $N_{-h_W+1}$, then there is no conserved quantity associated with $W$; if we can show that for every field $W$ of weight $h_W$, the corresponding state $\ket{V_1}$ is not a total derivative, then we have shown there is no local conserved charge of spin $h_W-1$.

\bibliographystyle{utphys}
\bibliography{refs}

\end{document}